\documentclass[10pt, a4paper, twocolumn]{article}

\usepackage[english]{babel} % English language hyphenation

\usepackage{microtype} % Better typography

\usepackage{amsmath,amsfonts,amsthm} % Math packages for equations

\usepackage[svgnames]{xcolor} % Enabling colors by their 'svgnames'

\usepackage[hang, small, labelfont=bf, up, textfont=it]{caption} % Custom captions under/above tables and figures

\usepackage{booktabs} % Horizontal rules in tables

\usepackage{lastpage} % Used to determine the number of pages in the document (for "Page X of Total")

\usepackage{graphicx} % Required for adding images

\usepackage{enumitem} % Required for customising lists
\setlist{noitemsep} % Remove spacing between bullet/numbered list elements

\usepackage{sectsty} % Enables custom section titles
\allsectionsfont{\usefont{OT1}{phv}{b}{n}} % Change the font of all section commands (Helvetica)

\usepackage{geometry} % Required for adjusting page dimensions

\usepackage[T1]{fontenc} % Output font encoding for international characters
\usepackage[utf8]{inputenc} % Required for inputting international characters

\usepackage{XCharter} % Use the XCharter font

\usepackage{fancyhdr} % Needed to define custom headers/footers
\fancypagestyle{firstpage}{ % Page style for the first page with the title
	\fancyhf{}
}

\newcommand{\authorstyle}[1]{{\large\usefont{OT1}{phv}{b}{n}\color{DarkRed}#1}} % Authors style (Helvetica)

\newcommand{\institution}[1]{{\footnotesize\usefont{OT1}{phv}{m}{sl}\color{Black}#1}} % Institutions style (Helvetica)

\usepackage{titling} % Allows custom title configuration

\newcommand{\HorRule}{\color{DarkGoldenrod}\rule{\linewidth}{1pt}} % Defines the gold horizontal rule around the title

\pretitle{
	\vspace{-30pt} % Move the entire title section up
	\HorRule\vspace{10pt} % Horizontal rule before the title
	\fontsize{32}{36}\usefont{OT1}{phv}{b}{n}\selectfont % Helvetica
	\color{DarkRed} % Text colour for the title and author(s)
}

\posttitle{\par\vskip 15pt} % Whitespace under the title

\preauthor{} % Anything that will appear before \author is printed

\postauthor{ % Anything that will appear after \author is printed
	\vspace{10pt} % Space before the rule
	\par\HorRule % Horizontal rule after the title
	\vspace{20pt} % Space after the title section
}

\usepackage{lettrine} % Package to accentuate the first letter of the text (lettrine)
\usepackage{fix-cm}	% Fixes the height of the lettrine

\newcommand{\initial}[1]{ % Defines the command and style for the lettrine
	\lettrine[lines=3,findent=4pt,nindent=0pt]{% Lettrine takes up 3 lines, the text to the right of it is indented 4pt and further indenting of lines 2+ is stopped
		\color{DarkGoldenrod}% Lettrine colour
		{#1}% The letter
	}{}%
}

\usepackage{xstring} % Required for string manipulation

\newcommand{\lettrineabstract}[1]{
	\StrLeft{#1}{1}[\firstletter] % Capture the first letter of the abstract for the lettrine
	\initial{\firstletter}\textbf{\StrGobbleLeft{#1}{1}} % Print the abstract with the first letter as a lettrine and the rest in bold
}

\usepackage{hyperref, booktabs, siunitx}
\DeclareSIUnit{\muas}{\mu\text{as}}
\DeclareSIUnit{\jy}{\text{Jy}}
\usepackage{amsmath, amssymb, mathrsfs, dsfont, nicefrac, mathtools}
\usepackage{cleveref}
\usepackage{graphicx}
\usepackage{subcaption}
\usepackage[mode=buildmissing]{standalone}
\usepackage{csquotes}
\usepackage{xr}
\usepackage{tikz}
\usetikzlibrary{graphs, positioning, shapes.geometric, arrows.meta, shapes.multipart}
\usepackage{pgfplots}
\pgfplotsset{compat=1.14}

\usepackage[backend=biber,style=nature]{biblatex}
\bibliography{sn-mathphys}

\title{Variable structures in M87* from space, time and frequency resolved interferometry}
\author{
	\authorstyle{Philipp Arras\textsuperscript{1,2}, Philipp Frank\textsuperscript{1,3}, Philipp Haim\textsuperscript{1}, Jakob Knollmüller\textsuperscript{1,2}, Reimar Leike\textsuperscript{1}, Martin Reinecke\textsuperscript{1}, and Torsten Enßlin\textsuperscript{1}} % Authors
	\newline\newline % Space before institutions
	\textsuperscript{1}\institution{Max-Planck Institut f\"ur Astrophysik, Karl-Schwarzschild-Str.~1, 85748 Garching, Germany}\\
	\textsuperscript{2}\institution{Technische Universit\"at M\"unchen, Boltzmannstr.~3, 85748 Garching, Germany}\\
	\textsuperscript{3}\institution{Ludwig-Maximilians-Universit\"at M\"unchen, Geschwister-Scholl-Platz~1, 80539 M\"unchen, Germany}
}

\date{January 03, 2022}
\begin{document}
\maketitle

\thispagestyle{firstpage} % Apply the page style for the first page (no headers and footers)

\lettrineabstract{Observing the dynamics of compact astrophysical objects provides insights into their inner workings, thereby probing physics under extreme conditions.
The immediate vicinity of an active supermassive black hole with its event horizon, photon ring, accretion disk, and relativistic jets is a perfect place to study general relativity and magneto-hydrodynamics.
The observations of M87* with {Very Long Baseline Interferometry} (VLBI) by the {Event Horizon Telescope} (EHT, \cite{ehti, ehtii, ehtiii, ehtiv, ehtv, ehtvi}) allows to investigate its dynamical processes on time scales of days.
Compared to regular radio interferometers, VLBI networks typically have fewer antennas and low signal to noise ratios (SNRs).
Furthermore, the source is variable, prohibiting integration over time to improve SNR.
Here, we present an imaging algorithm \cite{vlbiresolve, zenodo_software} that copes with the data scarcity and temporal evolution, while providing uncertainty quantification.
Our algorithm views the imaging task as a Bayesian inference problem of a time-varying brightness, exploits the correlation structure in time, and reconstructs a ${2+1+1}$ dimensional time-variable and spectrally resolved image at once.
We apply this method to the EHT observation of M87* \cite{ehtdata} and validate our approach on synthetic data.
The time- and frequency-resolved reconstruction of M87* confirms variable structures on the emission ring.
The reconstruction indicates extended and time-variable emission structures outside the ring itself.}

To address the imaging challenge of time-resolved VLBI data, we employ Bayesian inference. %
In particular, we adopt the formalism of \emph{information field theory} (IFT) \cite{ensslin18} for the inference of field-like quantities such as the sky brightness. 
IFT combines the measurement data and any included prior information into a consistent sky brightness reconstruction and propagates the remaining uncertainties into all final science results. %
Assuming limited spatial, frequency, and temporal variations, we can work with sparsely sampled data, such as the 2017 EHT observation of M87*. %

A related method based on a Gaussian Markov model was proposed by \cite{bouman2017} and another approach based on constraining information distances between time frames was proposed by \cite{Johnson_2017}. %
These methods impose fixed correlations in space or time, whereas our approach adapts flexibly to the demands of the data. %
We also enforce strict positivity of the brightness and instead of maximizing the posterior probability, we perform a variational approximation, taking uncertainty correlations between all model parameters into account. %

Interferometers sparsely probe the Fourier components of the source brightness distribution. %
The measured Fourier modes, called visibilities, are determined by the orientation and distance of antenna pairs, while the Earth's rotation helps to partly fill in the gaps by moving these projected baselines within the source plane. %
Since the source is time-variable and we aim at a time-dependent reconstruction, the measurement data have to be subdivided into multiple separate image frames along the temporal axis, leading to an extremely sparse Fourier space coverage in every frame.
%this coverage in Fourier coordinates is extremely sparse, as measurements at different times are looking at a changed source and need to be represented by separate image frames. %
In the case of the EHT observation of M87*, data were taken during four 8-hour cycles spread throughout seven days.
All missing image information needs to be restored by the imaging algorithm, exploiting implicit and explicit assumptions about the source structure.  %

Physical sources, including M87*, evolve continuously in time.
Images of these sources separated by time intervals that are short compared to the evolutionary time scale are thus expected to be strongly correlated.
Imposing these expected correlations during the image reconstruction process can inform image degrees of freedom (DOFs) that are not directly constrained by the data.

In radio interferometric imaging, spatial correlations can be enforced by convolving the image with a kernel, either during imaging, as part of the regularisation, or as a post-processing step. %
In our algorithm, we use a kernel as part of a forward model, where an initially uncorrelated image is convolved with the kernel to generate a proposal for the logarithmic sky brightness distribution, which is later adjusted to fit the data.
The specific structure of such a kernel can have substantial impact on the image reconstruction. %
We infer this kernel in a non-parametric fashion simultaneously with the image.
This substantially reduces the risk of biasing the result by choosing an inappropriate kernel, at the cost of introducing redundancies between DOFs of the convolution kernel and those of the pre-convolution image.

\emph{Metric Gaussian Variational Inference} (MGVI) is a Bayesian inference algorithm that is capable of tracking uncertainty correlations between all involved DOFs, which is crucial for models with redundancies, while having memory requirements that grow only linearly with the number of DOFs \cite{mgvi}.
It represents uncertainty correlation matrices implicitly without the need for an explicit storage of their entries and provides uncertainty quantification of the final reconstruction in terms of samples drawn from an approximate Bayesian posterior distribution, with a moderate level of approximation.  %
Compared to methods that provide a best-fit reconstruction, our approach provides a probability distribution, capturing uncertainty. %

A limitation of the Gaussian approximation is its uni-modality, as the posterior distribution is multi-modal \cite{bouman_normalizing_flows_2020}.
Representing multi-modal posteriors in high dimensions is hard if not infeasible.
Therefore, our results describe a typical mode of this distribution, taking the probability mass into account.

MGVI is the central inference engine of the Python package \emph{Numerical Information Field Theory} \cite[NIFTy]{nifty5,niftycode,zenodo_software}, which we use to implement our imaging algorithm, as it permits the flexible implementation of hierarchical Bayesian models.
NIFTy turns a forward model into the corresponding backward inference of the model parameters by means of automatic differentiation and MGVI. %
For time-resolved VLBI imaging, we therefore need to define a data model that encodes all relevant physical knowledge of the measurement process and the brightness distribution of the sky. %

This forward model describes in one part the sky brightness, and in another part the measurement process.
For the sky brightness, we require strictly positive structures with characteristic correlations in space, time, and frequency. %
These brightness fluctuations can vary exponentially over linear distances and time intervals,
which is represented by a log-normal prior with a Gaussian process kernel. %
The correlation structure of this process is assumed to be statistically homogeneous and isotropic for space, time, and frequency individually and decoupled for each sub-domain.
Consequently the correlations are represented by a direct outer product of rotationally symmetric convolution kernels, or equivalently by a product of one-dimensional, isotropic power spectra in the Fourier domain. %
We assume the power spectra to be close to power laws with deviations modelled as an integrated Wiener processes on a double logarithmic scale \cite{integratedwienerprocess}.
The DOFs, which finally determine the spatio-temporal correlation kernel, are inferred by MGVI alongside the sky brightness distribution. %
While the adopted model can only describe homogeneous and isotropic correlations, this symmetry is broken for the sky image itself by the data, which in general enforce heterogeneous and anisotropic structures.

The EHT collaboration has published data averaged down to two frequency bands at \SIlist{227;229}{\giga\hertz}.
Therefore, we employ a simplified model for the frequency axis:
We reconstruct two separate, but correlated images for these bands, with a priori assumed log-normal deviation on the \SI{1}{\percent} level, which amounts to spectral indices of $\pm 1$ within one standard deviation. %
Our algorithm does not constrain the absolute flux of the two channels.
Thus, we can recover the relative spectral index changes throughout the source but not the absolute ones.
A detailed description of the sky model is outlined in the methods section. %

We further require an accurate model of the instrument response. %
Just as the prior model is informed by our physical knowledge of the source, the instrument model is informed by our knowledge of the instrument. %
We consider two sources of measurement noise that cause the observed visibilities to differ from the perfect sky visibilities, %
the first being additive Gaussian thermal noise, whose magnitude is provided by the EHT collaboration in the data set.
The other component consists of multiplicative, systematic measurement errors, which are mainly caused by antenna-based effects, e.g.\ differences in the measurement equipment, atmospheric phase shift, and absorption of the incoming electromagnetic waves.  %
This source of errors can be conveniently eliminated by basing the model on derived quantities (closure amplitudes and phases), which are not affected by it. %
All those effects can be summarized in one complex, possibly time-variable, number per telescope, containing the antenna gain factors and antenna phases. %

For VLBI on \si{\muas}-scale, these effects can be prohibitively large.
Fortunately, certain combinations of visibilities are invariant under antenna-based systematic effects, so called closure-phases and -amplitudes  \cite{originalclosurephases}. %
These quantities serve as the data for our reconstruction (for details refer to Methods section). %

We apply this method to the EHT data of the super-massive black hole M87*.
With a shadow of the size of approximately four light days and reported superluminal proper motions of $6c$ \cite{biretta1999hubble}, its immediate vicinity is expected to be highly dynamic and subject to change on a time scale of days. %
The exceptional angular resolution of the EHT allowed for the first time to image the shadow of this super-massive black hole directly and to confirm its variability on horizon scale.

\begin{figure*}
  \centering
  \includegraphics{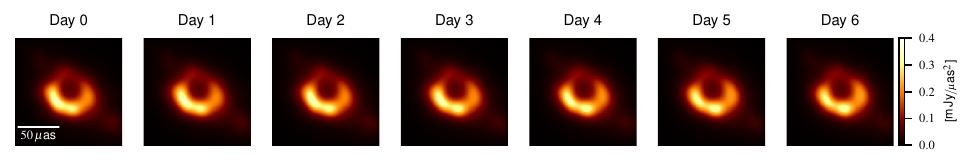}
  \vspace{-0.8em}

  \hspace*{1cm}%
  \includegraphics{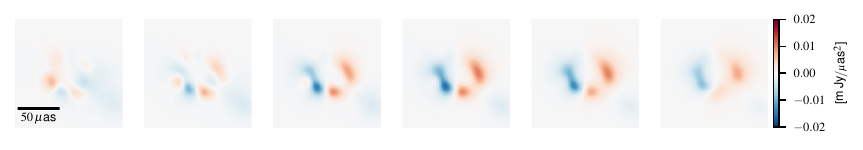}
  \vspace{-0.8em}

  \hspace*{1cm}%
  \includegraphics{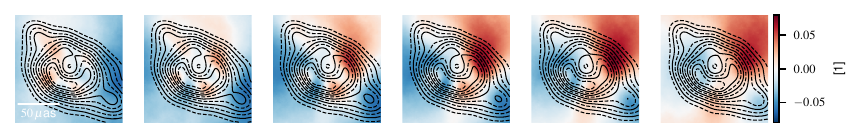}
  \caption[Temporal evolution of the brightness distribution]{
    Temporal evolution of the brightness distribution. %
    All figures are constrained to half the reconstructed field of view. %
    The first row shows time frames of the image cube, one for each day. %
    The second row visualises the brightness for day $N+1$ minus day $N$. %
    Red and blue visualises increasing and decreasing brightness over time, respectively. %
    The third row visualises the relative difference in brightness over time. %
    The over-plotted contour lines show brightness in multiplicative steps of $\nicefrac{1}{\sqrt{2}}$ and start at the maximum of the posterior mean of our reconstruction.
    The solid lines correspond to factors of powers of two from the maximum.
  }
  \label{fig:evolution}
\end{figure*}

In this letter, we present a time- and frequency-resolved reconstruction of the shadow of M87* over the entire observational cycle of seven days, utilizing correlation in all four dimensions (see \cref{fig:evolution}).
The closure quantities do not contain information on the total flux and the absolute position of the source.
Therefore, we normalize our results such that the flux in the entire ring is constant in time and agrees with the results of the EHT collaboration for the first frame of our reconstruction.
To achieve an alignment of the source even in the absence of absolute position information we start the inference with the data of only the first two observation days and continue with all data until convergence.

\begin{figure*}
  \centering
  \includegraphics[width=\textwidth]{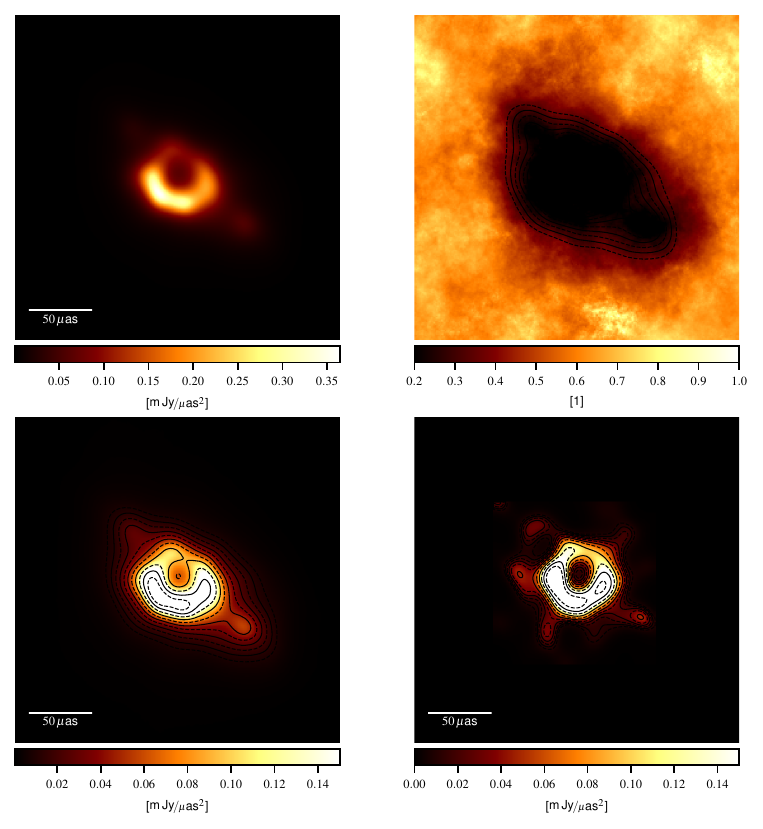}
  \caption[Brightness distribution on the first day]{
    Brightness distribution on the first day.
    The top row shows the reconstructed mean and relative error. %
    		Note that the small-scale structure in regions with high uncertainty in the error map is an artefact of the limited number of samples. %
    The bottom left shows a saturated plot of the approximate posterior mean, revealing the emission zones outside the ring. %
    The bottom right shows the result of the EHT-imaging pipeline in comparison, saturated to the same scale and with overplotted contour lines. %
    The over-plotted contour lines show brightness in multiplicative steps of $\nicefrac{1}{\sqrt{2}}$ and start at the maximum of the posterior mean of our reconstruction.
    The solid lines correspond to factors of powers of two from the maximum.
  }
  \label{fig:saturated}
\end{figure*}

\Cref{fig:saturated} displays the frequency-averaged sample mean image for the first observing day together with its pixel-wise uncertainty.
In full agreement with the EHT result, our image shows an emission ring that is brighter on its southern part, most likely due to relativistic beaming effects. %
Additionally, we obtain two faint extended structures, positioned opposite to each other along the south-western and north-eastern direction.
They do not have the shape of typical VLBI-imaging artefacts, i.e.\ they are not faint copies of the source itself, and similar structures do not appear in any of our validation examples.
We conclude that these structures are either of physical origin or due to unmodelled effects of the measurement in our algorithm.
These include baseline-based calibration artefacts such as polarization leakage \cite{ehtiii}, and extended emission outside the field of view. 
The latter likely has only a small effect, as we do not use closures that contain intra-site baselines, and all others should be insensitive to the large-scale jet emission \cite{ehtiv}.
The detection of additional significant source features, compared to the results by the EHT collaboration, is enabled by the usage of the data of all four observation days at once and thereby partially integrating the information.

\begin{figure*}
  \centering
  \includegraphics[width=\textwidth]{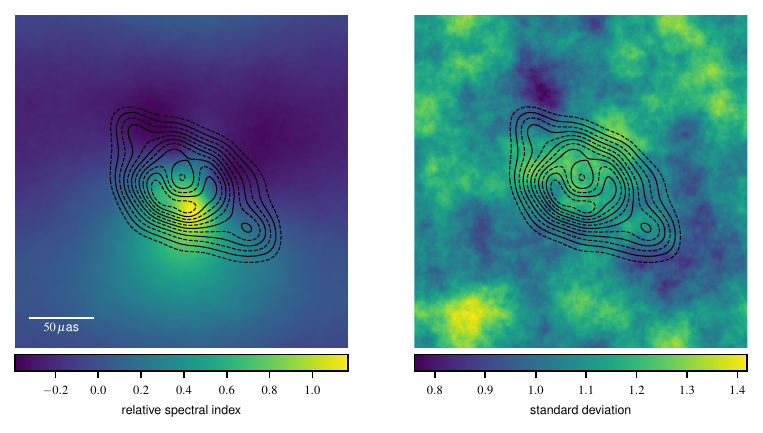} \\
   \includegraphics[width=\textwidth]{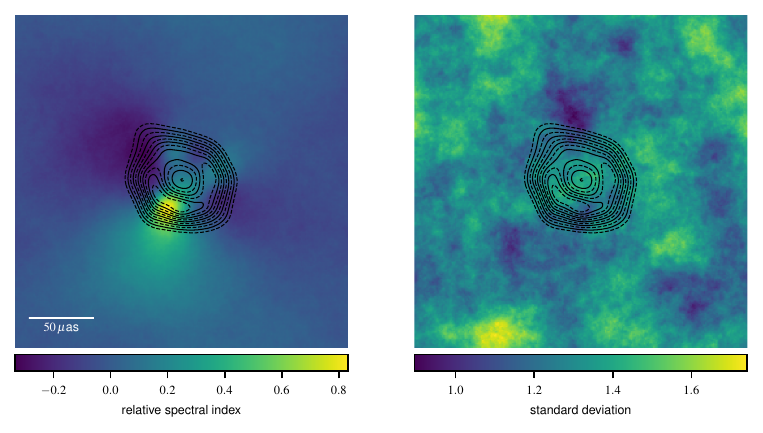}
  \caption[Relative spectral index]{Relative spectral index.
	Mean and pixel-wise uncertainty of the relative spectral index, as calculated from the \SIlist{227;229}{\giga\hertz} channels for M87* (top) and the \texttt{eht-crescent} example (bottom).
	}
  \label{fig:spectralindex}
\end{figure*}

Since our reconstruction is based on closure quantities that are not sensitive to absolute flux, the absolute spectral dependency is not constrained.
Still, the relative spectral index variations w.r.t.\ an overall spectrum can be explored (see top row of \cref{fig:spectralindex}).
The map exhibits a higher relative spectral index in the southern portion of the ring which coincides with its brightest emission spot.
However, the uncertainty map indicates that this feature is not significant and similar features falsely appear in the validation (see bottom row of \cref{fig:spectralindex}).
Therefore, we do not report any significant structures in the spectral behaviour of M87* and continue our analysis with frequency-averaged time frames.

\begin{figure*}
  \centering
  \includegraphics[width=\textwidth]{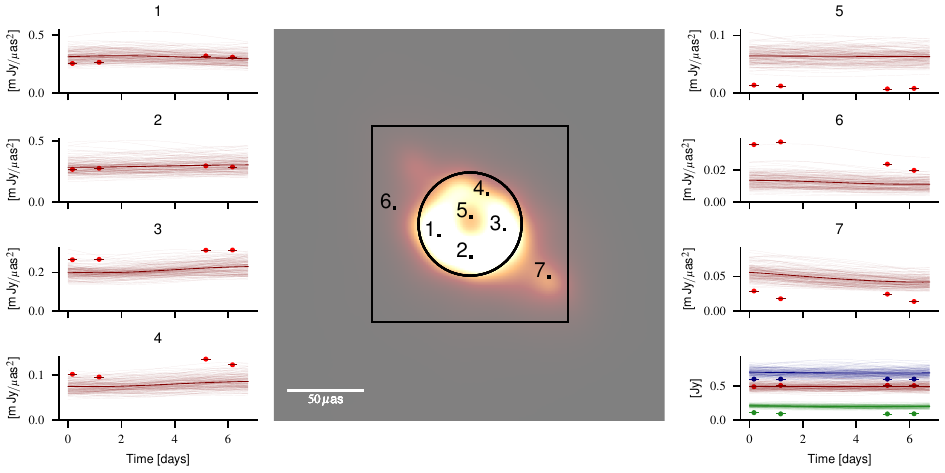}
  \caption[Temporal evolution at selected locations]{
    Temporal evolution at selected locations.
    The brightness and flux for approximate posterior samples and their ensemble mean at specific sky locations and areas as indicated in the central panel. %
    The peripheral panels show brightness and flux values of samples (thin lines) and their mean (thick lines). %
    Of those, the bottom right one displays the flux inside (red) and outside the circle (green), as well as the sum of the two (blue). %
    For comparability, only brightness within the field of view of the EHT collaboration image, indicated by the black box in the central plot, is integrated. %
    The remaining panels give the local brightness for the different locations labelled by numbers in the central panel. %
    The single-day results from EHT-imaging are indicated as points.
  }
  \label{fig:timeseries}
\end{figure*}

The sky brightness for each day of the observation together with the absolute and relative differences between adjacent days is displayed in \cref{fig:evolution}.
We report mild temporal brightness changes of up to \SI{6}{\percent} per day, in particular within the western and southern parts of the ring, validating the observations made by \cite{ehtiv}.
\Cref{fig:timeseries} shows the detailed temporal evolution of a selected number of locations and areas.
Our method consistently interpolates in between observations.
Supplementary Video 1 also demonstrates the continuous evolution.
In several locations our reconstruction agrees with the EHT's imaging results, whereas others clearly deviate.
Especially at location~7, which corresponds to the extended structure in the south-western direction, the brightness decreases by about \SI{5}{\percent} between adjacent days throughout the entire observation.
This hints at a real and non-trivial temporal evolution.

\begin{table*}
  \centering
  
\begin{tabular}{lccccc}
\toprule
& $d\, (\mu\text{as})$ & $w\, (\mu\text{as})$ & $\eta\, (^\circ )$ & $A$ & $f_C$\\\midrule

\multicolumn{6}{l}{\textsc{difmap}}\\
April 5 & $37.2 \pm 2.4$ & $ 28.2 \pm 2.9 $ & $163.8 \pm 6.5$& $0.21 \pm 0.03$& $0.5$\\
April 6 & $40.1 \pm 7.4$ & $ 28.6 \pm 3.0$& $162.1 \pm 9.7$& $0.24 \pm 0.08$& $0.4$\\
April 10 & $40.2 \pm 1.7$ & $ 27.5 \pm 3.1$& $175.8 \pm 9.8$& $0.20 \pm 0.04$& $0.4$\\
April 11 & $40.7 \pm 2.6$ & $ 29.0 \pm 3.0$& $173.3 \pm 4.8$& $0.23 \pm 0.04$& $0.5$\\\midrule

\multicolumn{1}{l}{\textsc{eht-imaging}}\\
April 5 & $39.3 \pm 1.6$ & $ 16.2 \pm 2.0 $ & $148.3 \pm 4.8$& $0.25 \pm 0.02$& $0.08$\\
April 6 & $39.6 \pm 1.8$ & $ 16.2 \pm 1.7$& $151.1 \pm 8.6$& $0.25 \pm 0.02$& $0.06$\\
April 10 & $40.7 \pm 1.6$ & $ 15.7 \pm 2.0$& $171.2 \pm 6.9$& $0.23 \pm 0.03$& $0.04$\\
April 11 & $41.0 \pm 1.4$ & $ 15.5\pm 1.8$& $168.0 \pm 6.9$& $0.20 \pm 0.02$& $0.04$\\\midrule

\multicolumn{6}{l}{\textsc{smili}}\\
April 5 & $40.5 \pm 1.9$ & $16.1 \pm 2.1$ & $154.2 \pm 7.1$& $0.27 \pm 0.03$& $7 \times 10^{-5} $\\
April 6 & $40.9 \pm 2.4$ & $16.1 \pm 2.1$& $151.7 \pm 8.2$& $0.25 \pm 0.02$& $2 \times 10^{-4}  $\\
April 10 & $42.0 \pm 1.8$ & $15.7 \pm 2.4$& $170.6 \pm 5.5$& $0.21 \pm 0.03$& $4 \times 10^{-6}  $\\
April 11 & $42.3 \pm 1.6$ & $15.6 \pm 2.2$& $167.6 \pm 2.8$& $0.22 \pm 0.03$& $6 \times 10^{-6}  $\\\midrule
\multicolumn{6}{l}{\textsc{Our method (uncertainty as per \cite[Table 7]{ehtiv})}}\\
April 5 & $44.4 \pm 3.4$ & $23.2 \pm 5.2$ & $164.9 \pm 9.5$ & $0.26 \pm 0.04$ & $0.365 $ \\
April 6 & $44.4 \pm 2.9$ & $23.3 \pm 5.4$ & $161.7 \pm 5.6$ & $0.24 \pm 0.04$ & $0.374 $ \\
April 10 & $44.8 \pm 2.8$ & $23.0 \pm 5.0$ & $176.7 \pm 9.8$ & $0.22 \pm 0.03$ & $0.374 $ \\
April 11 & $44.6 \pm 2.8$ & $22.8 \pm 4.8$ & $180.1 \pm 10.4$ & $0.22 \pm 0.03$ & $0.372 $ \\\midrule
\multicolumn{6}{l}{\textsc{Our method (sample uncertainty)}}\\
April 5 & $44.1 \pm 1.2$ & $23.1 \pm 2.4$ & $163.9 \pm 5.0$ & $0.25 \pm 0.03$ & $0.377 \pm 0.081$\\
April 6 & $44.0 \pm 1.2$ & $22.9 \pm 2.4$ & $161.9 \pm 6.0$ & $0.24 \pm 0.03$ & $0.385 \pm 0.085$\\
April 10 & $44.6 \pm 1.2$ & $22.9 \pm 2.5$ & $176.2 \pm 6.5$ & $0.22 \pm 0.03$ & $0.383 \pm 0.089$\\
April 11 & $44.6 \pm 1.2$ & $23.0 \pm 2.6$ & $179.8 \pm 6.2$ & $0.22 \pm 0.03$ & $0.383 \pm 0.090$\\\bottomrule
\end{tabular}

  \caption[Extracted crescent parameters for M87*]{
    Extracted crescent parameters for M87*.
    A comparison of diameter $d$, width $w$, orientation angle $\eta$, asymmetry $A$ and floor-to-ring contrast ratio $f_C$ as defined by \cite[Table 7]{ehtiv} and computed for images published by the EHT collaboration (first three sections of table)  as well as for our reconstruction (last two sections). %
    Section four provides the result of the estimators and their standard deviations as defined by \cite{ehtiv} applied to our posterior mean. %
    Section five provides means and standard deviations based on processing our posterior samples individually through the estimators and by computing mean and 1-$\sigma$ standard deviations from these results. %
  }
  \label{tab:ringfits}
\end{table*}

Following the analysis of \cite{ehtiv}, we compute empirical characteristics of the asymmetric ring, i.e.\ diameter $d$, width $w$, orientation angle $\eta$, azimuthal brightness asymmetry $A$, and floor-to-ring contrast ratio $f_C$.
All findings are summarized in \cref{tab:ringfits} and compared to the results of the EHT collaboration \cite{ehtiv}:
We can confirm the stationary values for diameter $d$, width $w$, azimuthal brightness asymmetry $A$, and floor-to-ring contrast ratio $f_C$ during the seven days and a significant temporal evolution of the orientation angle $\eta$.
The latter might be caused by flickering of emission spots \cite{2020AA...634A..38N}.
We report a slightly larger diameter $d = \SI{45\pm 3}{\muas}$, which does not significantly deviate from the result published by the EHT Collaboration of $d=\SI{42 \pm 3}{\muas}$ \cite{ehti}.

\begin{figure*}
  \centering
  \includegraphics{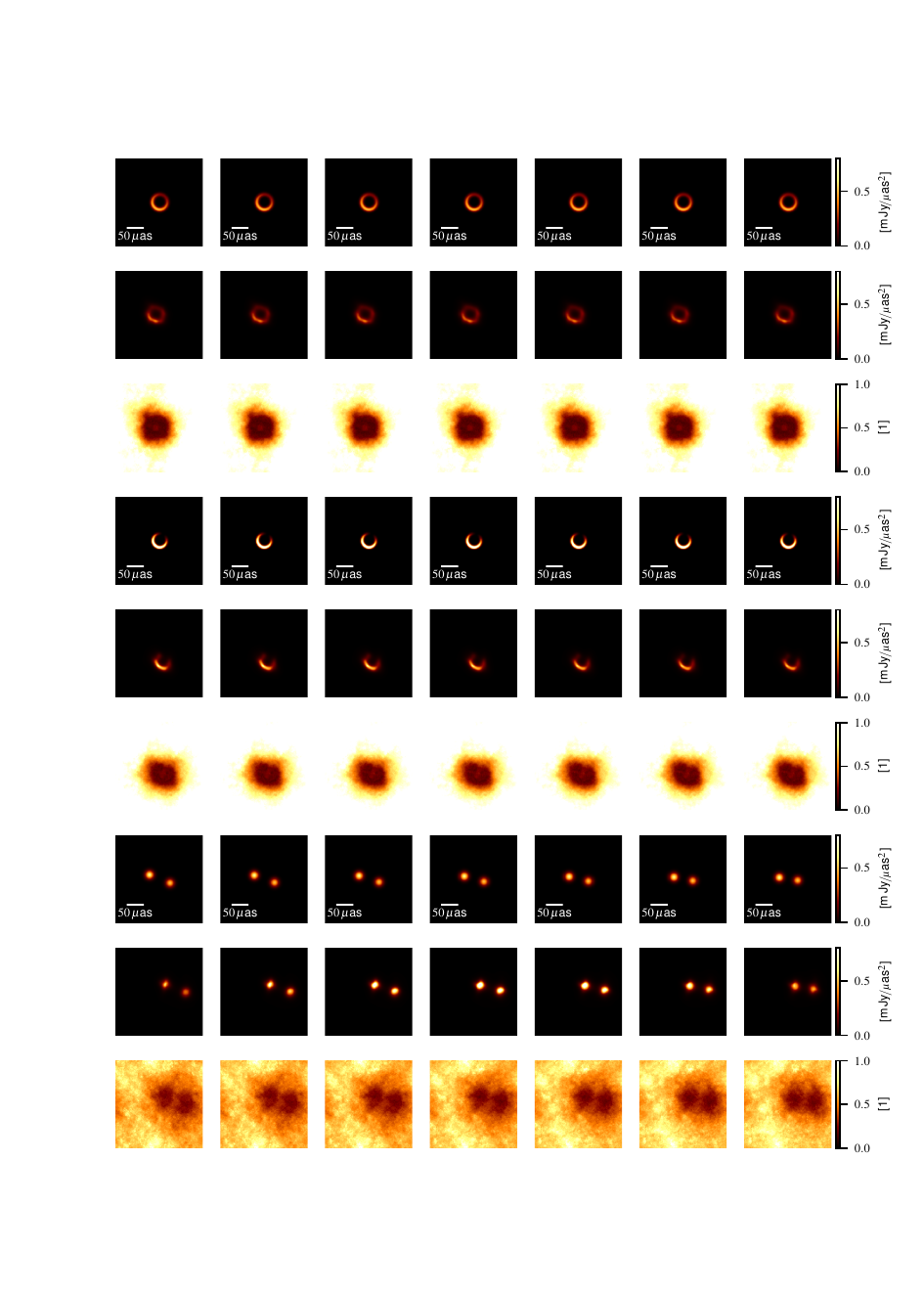}
  \caption[Validation on synthetic observations of time-variable sources]{
    Validation on synthetic observations of time-variable sources.
    In the figure, time goes from left to right showing slices through the image cube for the first time bin of each day.
    Different source models are shown from top to bottom: \texttt{eht-crescent}, \texttt{slim-crescent}, and \texttt{double-sources}.
    For each source the ground truth, the approximate posterior mean of the reconstruction, and the relative standard deviation, clipped to the interval $[0, 1]$, are displayed (from top to bottom).
    The central three columns show moments in time in which no data is available
    since data was taken only during the first and last two days of the week-long observation period.
  }
  \label{fig:validation}
\end{figure*}

\begin{figure*}
  \centering
  \includegraphics{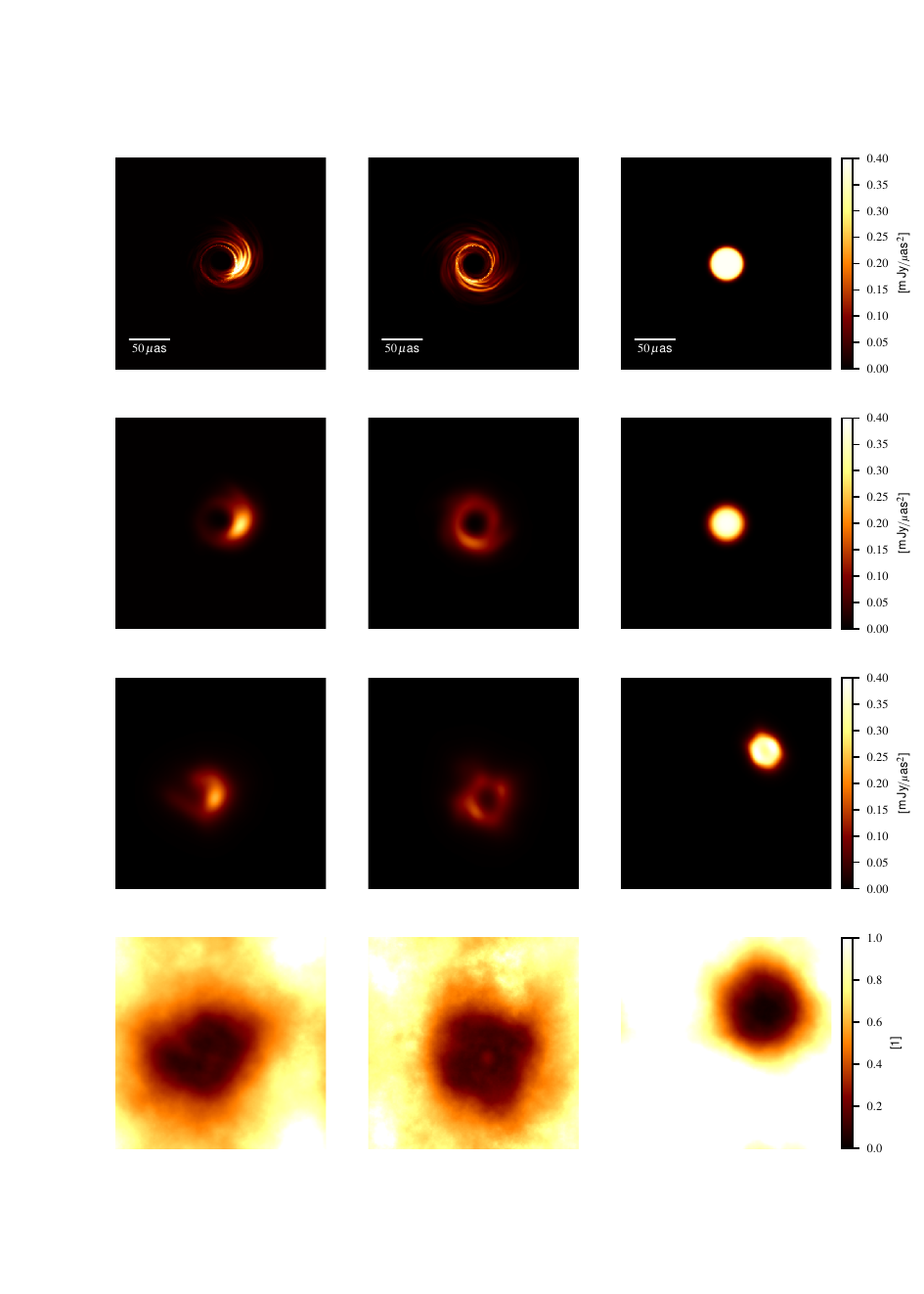}
  \caption[Validation for static sources]{
    Validation for static sources.
    We show two scenarios from the EHT imaging challenge and a uniform disk.
    The rows depict the ground truth, the smoothed ground truth, the  approximate posterior mean, and the relative standard deviation for our three static validation examples.
    The plots in the first three rows are normalized to their respective maximum, are not clipped, and the minimum of the colour bar is zero.
    In the last row the colour bar is clipped to the interval $[0, 1]$.
  }
  \label{fig:staticvalidation}
\end{figure*}

\begin{figure*}
    \centering
    \begin{subfigure}[b]{0.49\textwidth}
        \includegraphics[width=\textwidth]{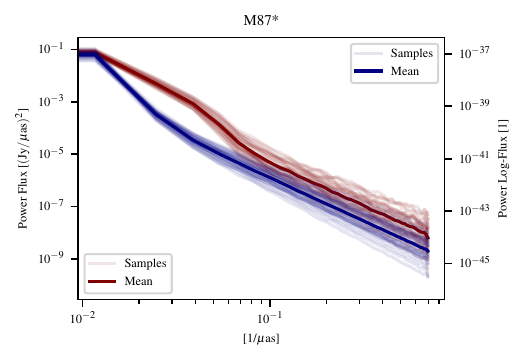}
        \includegraphics[width=\textwidth]{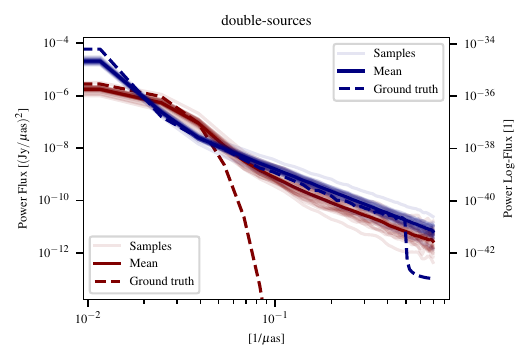}
        \includegraphics[width=\textwidth]{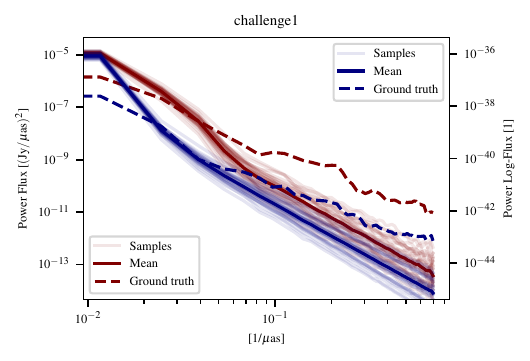}
    \end{subfigure}
    \begin{subfigure}[b]{0.49\textwidth}
        \includegraphics[width=\textwidth]{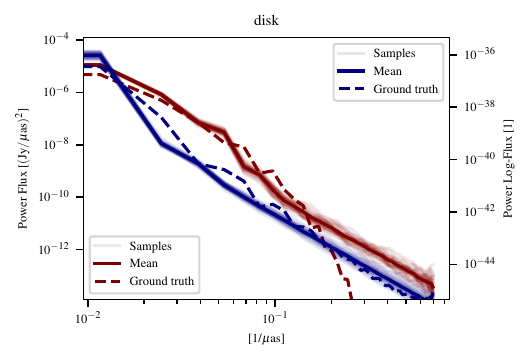}
        \includegraphics[width=\textwidth]{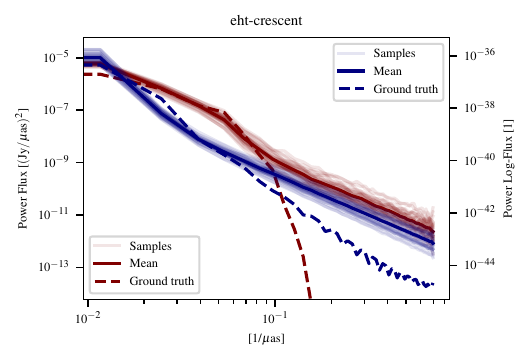}
        \includegraphics[width=\textwidth]{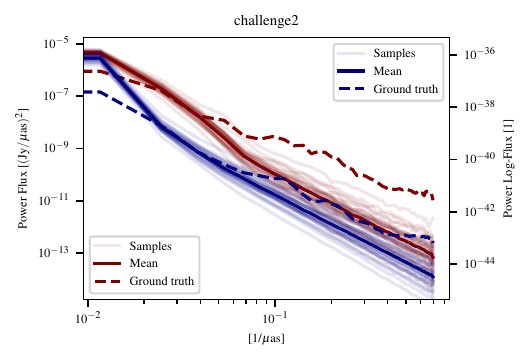}
    \end{subfigure}
    \caption[The reconstructed spatial correlation power spectra]{
        The spatial correlation power spectra of our reconstruction for the EHT-observation of M87* (top left panel) and five of our validation data sets. %
        The red curves show the power spectra of the reconstructed brightness. %
        The blue curves show the power spectra of the logarithmic brightness. %
        For the three validation sets, the corresponding power spectra of the ground truth are plotted as a dashed line. %
    }
    \label{fig:ps}
\end{figure*}

\begin{figure*}
  \centering
  \includegraphics[width=\textwidth]{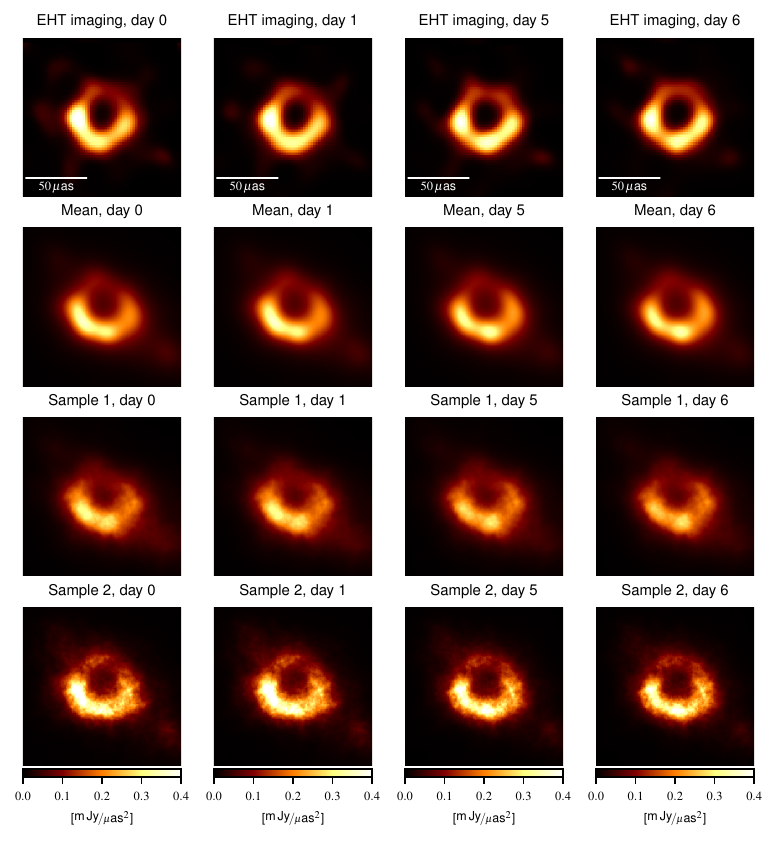}
  \caption[Comparison of our imaging result to that of the EHT-imaging pipeline]{
    Comparison of our imaging result to that of the EHT-imaging pipeline. %
    All panels have the same colour bar. %
    The columns label the four days for which observational data exist. %
    The first row shows snapshot images from the EHT-imaging pipeline for each of the 4 days. %
    The second row shows our mean reconstruction for the same time frame. %
    The third and fourth row each show one posterior sample from our imaging pipeline. %
  }
  \label{fig:comparison}
\end{figure*}

A collection of six validation examples has been assembled to assess accuracy and robustness of our method (\cref{fig:validation,fig:staticvalidation}).
\cref{fig:ps} shows spatial correlation spectra for our scientific and validation images.
\cref{fig:comparison} displays the results of the  imaging methods used by the EHT Collaboration together with our posterior mean and two samples for all observation periods.
The temporal evolution of several samples is illustrated in Supplementary Video 2.

In conclusion, we present and validate the first Bayesian imaging method that is capable of simultaneously reconstructing emission over spatial, temporal and spectral dimensions from closure quantities, utilizing correlation and quantifying uncertainties via posterior samples.
We provide the first independent confirmation of the overall morphology of the emission ring around M87* and an apparent evolution of its orientation as published by the EHT collaboration.
The frequency resolution allows us to obtain a relative spectral index map, together with an uncertainty estimation.
For the data set at hand, significant spectral features could not be found.
In addition to the emission ring, we resolve significant and potentially dynamic emission structures along the south-western and north-eastern direction. %
With future observations, our method may help to explore the intricate structure in the spatial, spectral, and temporal domain of M87* and other variable sources.
To achieve this, the model can be extended with inference of the prior spectral correlation structure.

\newcommand{\chitwovalue}{1.48}

\section*{Methods}

The reconstruction algorithm relies on Bayesian statistics.
Thus, it consists of three essential components: the likelihood, the prior, and an inference scheme.

The likelihood is a probabilistic description of the measurement process including details on the measurement device.
We choose to describe the measurement in terms of closure quantities that are invariant under antenna-based calibration effects.

The prior model captures all assumptions on the sky brightness distribution.
Here we assume positivity at all times, correlation along the temporal, spatial, and spectral direction, as well as the possibility of variations on an exponential scale.
This is implemented with the help of a Gaussian process prior of the logarithmic brightness distribution with unknown kernel.
Below, a non-parametric kernel model is derived that assumes a stochastic process along each dimension individually.

\begin{figure*}
  \centering
  \begin{tikzpicture}[
  x=10.5mm,
  % y=7mm,
  % node distance=30mm,
  % on grid,
  every node/.style={draw, anchor=west, thick},
  prior/.style={ellipse, draw, anchor=mid, dashed},
  arrow/.style={-{Stealth}, thick},
  normalleft/.style={draw, anchor=west, minimum width=49mm, align=center, minimum height=2em},
  normalright/.style={draw, anchor=east, minimum width=49mm, align=center, minimum height=2em},
  normalcenter/.style={draw, anchor=mid, minimum width=49mm, align=center, minimum height=2em},
  bigcenter/.style={draw, anchor=west, minimum width=105mm, align=center, minimum height=5em},
  ]

\node [prior, anchor=west] (prior_m) at (0, 3.5) {$\xi^{(i)}_m \sim \left.\mathcal{N}\left(\xi^{(i)}_m \right| 0, \mathds{1}\right)$};
\node [prior, anchor=east] (prior_eta) at (10, 3.5) {$\xi^{(i)}_\eta \sim \left.\mathcal{N}\left(\xi^{(i)}_\eta \right| 0, \mathds{1}\right)$};
\node [prior, anchor=east] (prior_w) at (-0.5, 0)  {$\xi^{(i)}_W \sim \left.\mathcal{N}\left(\xi^{(i)}_W \right| 0, \mathds{1}\right)$};
\node [prior, anchor=east] (amplitude_prior) at (-0.5, -3) {$\xi^{(i)}_a \sim \left.\mathcal{N}\left(\xi^{(i)}_a \right| 0, \mathds{1}\right)$};
\node [prior, anchor=east] (sky_prior) at (-0.5, -9) {$\xi_s \sim \left.\mathcal{N}\left(\xi_s \right| 0, \mathds{1}\right)$};
\node [prior, anchor=east] (global_scaling_prior) at (-0.5, -6) {$\xi_\alpha \sim \left.\mathcal{N}\left(\xi_\alpha \right| 0, \mathds{1}\right)$};

\node [bigcenter] (model_gamma) at (0, 0) {logarithmic coordinates $l = \log(|k|)$ \\ $\gamma^{(i)}(l) = m^{(i)} l + \eta^{(i)} \int_{l_0}^{l} \int_{l_0}^{l'} \xi^{(i)}_W(l'') \,dl' \,dl'' ,$};
\node [normalleft] (model_m) at (0, 2) {$m^{(i)} = \textcolor{teal}{\mu^{(i)}_{m}} + \textcolor{teal}{\sigma^{(i)}_{m}} \xi^{(i)}_{m}$};
\node [normalright] (model_eta) at (10, 2) {$\eta^{(i)}=e^{\textcolor{teal}{\mu^{(i)}_{\eta}} + \textcolor{teal}{\sigma^{(i)}_{\eta}} \xi^{(i)}_{\eta}}$};
\node [normalleft] (norm_fluct) at (0, -2) {$\widetilde{U}^{(i)} = \int_{k \neq 0} e^{2\,\gamma^{(i)}(|k|)} \, dk$};
\node [normalleft] (amplitude) at (0, -3) {$a^{(i)} = e^{\textcolor{teal}{\mu_a^{(i)}} + \textcolor{teal}{\sigma_a^{(i)}} \xi^{(i)}_a}$};
\node [normalright] (amplitudes_spec) at (10, -2.5) {Amplitudes \\ $A^{(i)}_{k k} = p^{(i)}(|k|) = a^{(i)}\, \frac{e^{\gamma^{(i)}(|k|)}}{\sqrt{\widetilde{U}^{(i)}}}$};

\node [normalleft, minimum width=0 mm] (volume) at (-.3, -4.5) {$V^{(i)} =  \int_{\Omega^{(i)}} \text{d}\Omega^{(i)}$};
\node [normalright] (normalized_amplitudes) at (10, -4.5) {$\widetilde{A}^{(i)} = \left(\frac{1}{V^{(i)}} \int_{\Omega^{(i)}} \left(F^{(i)}\right)^{-1} p^{(i)} \ \text{d}\Omega^{(i)} \right)^{-1} A^{(i)}$};

\node [normalleft] (global_scaling) at (0, -6) {$\alpha = e^{\textcolor{teal}{\mu_\alpha} + \textcolor{teal}{\sigma_\alpha} \xi_\alpha}$};
\node [normalright] (amplitude_operator) at (10, -6) {$A' = \alpha \bigotimes_{i\in\{x,t\}} \widetilde{A}^{(i)}$};

\node [normalcenter] (amplitude_operator_freq) at (5, -7.5) {$A = \begin{pmatrix}
	1 & \textcolor{teal}{\epsilon} \\
	1 & \textcolor{teal}{-\epsilon}
	\end{pmatrix} \bigotimes A'$};
\node [normalcenter] (log_sky) at (5, -9) {$\tau = A \xi_s$};
    \node [normalcenter, ultra thick] (sky) at (5, -10.5) {\textbf{Image} $s = \nicefrac{e^\tau}{\int \text{d}x\,e^\tau}$};

  \draw[arrow] (prior_m) -- (model_m);
  \draw[arrow] (prior_eta) -- (model_eta);
  \draw[arrow] (amplitude_prior) -- (amplitude);
  \draw[arrow] (global_scaling_prior) -- (global_scaling);

\draw[arrow] (model_m) -- (model_gamma);
\draw[arrow] (model_eta) -- (model_gamma);
\draw[arrow] (prior_w) -- (model_gamma);
\draw[arrow] (amplitude) -- (amplitudes_spec);
\draw[arrow] (norm_fluct) -- (amplitudes_spec);
\draw[arrow] (model_gamma) -- (norm_fluct);
\draw[arrow] (model_gamma) -- (amplitudes_spec);
\draw[arrow] (volume) -- (normalized_amplitudes);
\draw[arrow] (amplitudes_spec) -- (normalized_amplitudes);
\draw[arrow] (global_scaling) -- (amplitude_operator);
\draw[arrow] (normalized_amplitudes) -- (amplitude_operator);
\draw[arrow] (amplitude_operator) -- (amplitude_operator_freq);
\draw[arrow] (amplitude_operator_freq) -- (log_sky);
\draw[arrow] (log_sky) -- (sky);
\draw[arrow] (sky_prior) -- (log_sky);
\end{tikzpicture}
  \caption[Graphical structure of our model]{
    Graphical structure of our model.  
    The hierarchical model that was used as prior on the four-dimensional (frequency, time and space) image $s$, as described in the methods section.
    The round dashed nodes represent the inferred latent parameters, which are independent normal distributed a priori.
    The solid rectangular nodes represent computation steps.
    Arrows denote dependencies.
    All hyperparameters are marked in teal.
    The upper half of the diagram describes our non-parametric model of the power spectra in temporal and spatial domains.
    The lower half specifies how the four dimensional image is obtained from additional latent parameters and the power spectra.}
  \label{fig:hierarchicalmodel}
\end{figure*}

This constitutes a Bayesian inference problem that is approximately solved by applying Metric Gaussian Variational Inference (MGVI) as inference scheme.
This method requires a generative model formulation in which all model parameters are standard-normal distributed a priori.
The generative function defined below associates these with the physical quantities (see \cref{fig:hierarchicalmodel}).

We describe all implementation details and give the reasoning behind our choice of hyperparameters and the inference heuristic.
The method is validated on six simulated sources with a varying degree of dynamics, ranging from simple shapes to realistic black holes.
To demonstrate the effect of hyperparameter choices, we perform 100 reconstructions of both a synthetic example and M87* with randomized hyperparameters within a certain range.
All validation efforts show that the algorithm is able to reconstruct synthetic examples successfully and is stable under changes in the hyperparameters.

\subsection*{Likelihood}
The likelihood of the measured visibilities given the sky brightness distribution $s$ is computed independently for each time frame. %
The visibilities for all measured data points are assumed to follow the measurement equation in the flat sky approximation: %
\begin{align}
R(s)_{AB} &\coloneqq\int e^{-2\pi i\left(u_{AB}x+v_{AB}y\right)} s(x,y) \,dx\, dy \\
&\eqqcolon   e^{\rho_{AB}} e^{i \phi_{AB}} .
\end{align}
Here $AB$ runs through all ordered pairs of antennas $A$ and $B$ for all non-flagged baselines, $u_{AB}$ and $v_{AB}$ are the coordinates of the measured Fourier points, $s(x,y)$ is the sky brightness distribution as a function of sky angles $x$ and $y$, and $R$ is called measurement response.
The visibilities $R(s)_{AB}$ are complex numbers and we represent them in polar coordinates as phases $\phi_{AB}(s)\in \mathbb{R}$ and logarithmic amplitudes $\rho_{AB}(s)\in \mathbb{R}$, i.e.\ $R(s)_{AB}=\exp(\rho_{AB}(s)+i\,\phi_{AB}(s))$. %
We assume the thermal noise of the phase and logarithmic amplitude to be independently Gaussian distributed with covariance %
\begin{align}
N = \text{diag}\left( \frac{\sigma^2}{\vert d\vert^2}\right)\ ,
\end{align}
where $d$ is the reported visibility data and $\sigma$ is the reported thermal noise level. %
The operation $\text{diag}(x)$ denotes a diagonal matrix with $x$ on its diagonal. %
This is approximately valid for a signal-to-noise ratio larger than $5$ \cite{closure19}, which is true for most of our data. % quantifiend to be around one fourth being smaller than SNR 5

To avoid antenna based systematic effects, we compute closure quantities from these visibilities \cite{closure19}. %
Closure phases are obtained by combining a triplet of complex phases of visibilities via: %
\begin{align}
\left(\phi_{\text{cl}}\right)_{ABC} \coloneqq \phi_{AB} + \phi_{BC} + \phi_{CA}. \label{eq:define-closure-phase}
\end{align}
Closure amplitudes are formed by combining the logarithmic absolute value of four visibilities:
\begin{align}
\left(\rho_{\text{cl}}\right)_{ABCD} \coloneqq \rho_{AB} -\rho_{BC} + \rho_{CD} -\rho_{DA} .
\end{align}
These closure quantities are invariant under antenna based visibility transformations of the form
\begin{align}
R(s)_{AB} \rightarrow c_A c_B^* R(s)_{AB}
\end{align}
for all antennas and multiplicative calibration errors $c_A$ and $c_B$, where $*$ denotes the complex conjugate. %

Note that forming the closure phases is a linear operation on the complex phase, while forming the closure amplitudes is linear in the logarithmic absolute value. %
We can thus represent these operations using matrices: %
\begin{align}
\rho_{\text{cl}} = L \rho, \quad \phi_{\text{cl}}  = M \phi  .
\end{align}
The closure matrices $L$ and $M$ are sparse and contain in every row $\pm 1$ for visibilities associated with the closure, and zero elsewhere. %

The noise covariances $N_\rho$ and $N_\phi$ of the closure quantities are related to $N$ via: %
\begin{align}
N_\rho = \left<Ln(Ln)^\dagger\right>_{\mathcal{N}(n \vert 0, N)} = L N L^\dagger \quad &\text{and} \quad N_\phi = M N M^\dagger , \label{eq:noise-matrices}%
\end{align}
where $\dagger$ denotes the adjoint of the operator and $\mathcal{N}(n \vert 0, N)$ denotes a Gaussian distribution over $n$ with mean $0$ and covariance $N$.
The mixing introduced by applying $L$ and $M$ leads to non-diagonal noise covariance matrices of the closure quantities. %

For a given antenna setup (of five or more antennas), more closure quantities can be constructed than visibilities are available, and therefore they provide a redundant description of the data. %
For the logarithmic amplitudes $\rho$, we first construct all possible closure quantities and then map to a non-redundant set using the eigen-decomposition of $N_\rho$. %
Specifically, we construct a unitary transformation $U_\rho$ where each column of the matrix is an eigenvector corresponding to a non-zero eigenvalue of $N_\rho$. %
This transformation provides a map from the space of all possible closure amplitudes to the space of maximal non-redundant sets, with the additional property that the transformed noise covariance becomes diagonal. %
Specifically
\begin{equation}
U_\rho N_\rho U_\rho^\dagger = \Lambda_\rho \ ,
\end{equation}
where $\Lambda_\rho$ denotes a diagonal matrix with the non-zero eigenvalues of $N_\rho$ on its diagonal. %
We can combine $L$ and $U_\rho$ to form an operation that maps from the logarithmic amplitudes of visibilities $\rho$ directly to the space of non-redundant closure amplitudes $\varrho$ via
\begin{equation}
\varrho = U_\rho \rho_{\text{cl}} = U_\rho L \rho \ ,
\end{equation}
and use it to compute the observed, non-redundant closure amplitude $\varrho_d$ from the published visibility data $d = \exp({\rho_d + i\, \phi_d})$.

The resulting likelihood for closure amplitudes reads
\begin{equation}
\mathcal{P}( \varrho_d | \varrho, L, N) = \mathcal{N}(\varrho_d \vert \varrho, \Lambda_\rho) \ .
\end{equation}

Closure phases are constructed differently to avoid problems induced by phase wraps.
Adding or subtracting $2\pi$ from a phase does not change the result, and we need to preserve this symmetry in our algorithm.
We thus can only add integer multiples of phases such as \cref{eq:define-closure-phase} and this prohibits using a direct matrix decomposition to find a maximal non-redundant closure set.

We build the closure sets to be used in the imaging with the help of a greedy algorithm that processes closure phases in the order of decreasing signal-to-noise ratio, as defined by the inverse of the diagonal of $N_\phi$ (\cref{eq:noise-matrices}). %
The algorithm collects closure sets into $M$ until $\text{rank}(M) = \text{dim}(\phi)$ ensuring that $\phi_\text{cl}$ consists of a maximal non-redundant set. %
In principle, all maximal non-redundant closure sets are equivalent as long as one takes the non-diagonal noise covariance into account.
The concrete choice might have a minor impact for our approximation of the closure phase likelihood.

Within our closure set, we can decompose the noise covariance $N_\phi$ into a unitary matrix $U_\phi$ and its eigenvalues $\Lambda_\phi$.
Instead of working with the phases $\phi_{\text{cl}}$ directly, we use their positions on the complex unit circle $e^{i \phi_{\text{cl}}}$ to define
\begin{equation}
\varphi = U_\phi \ e^{i \phi_{\text{cl}}} = U_\phi \ e^{i M \phi} \ .
\end{equation}
This mitigates the problem of phase wraps at the price of approximating the corresponding covariance. %
This approximation yields errors below the \SI{1}{\percent} level if the signal-to-noise ratio is larger than $10$. %
Most of the data points are above that threshold, and the error decreases quadratically with increasing signal-to-noise ratio. %
Since data with the lowest standard deviation are also the most informative, we believe the impact of the approximation on the reconstruction to be negligible.

Given the closure phases on the unit circle $\varphi$, the corresponding phase likelihood can be written as
\begin{equation}
\mathcal{P}( \varphi_d | \varphi, L, N) = \mathcal{N}(\varphi_d \vert \varphi, \Lambda_\phi) \ , \label{eq:phase-noise}
\end{equation}
where $\varphi_d = U_\phi \ e^{i M \phi_d}$. %
Note that \cref{eq:phase-noise} is a Gaussian distribution on complex numbers with the probability density function as
\begin{align}
\mathcal{N}(x|y,X) = \vert 4\pi X \vert^{-\frac{1}{2}}\text{exp}\left(-\frac{1}{2}(x-y)^\dagger X^{-1}(x-y)\right)\ ,
\end{align}
and Hermitian covariance $X$.
Complex and real Gaussian distributions only differ in their normalization constant.
We do not distinguish between them explicitly, as the normalization is irrelevant for our variational approach. %

\subsection*{Modelling the sky brightness}
The sky brightness distribution $s_{x t \nu}$ is defined within a fixed field of view $\Omega_x \subset \mathbb R^2$, a time interval $\Omega_t = [0, \bar{t}]$, and frequency range $\Omega_\nu \subset \mathbb R$, which renders it to be a field defined in space, time, and frequency. %
We assume $s$ to be a priori log-normal distributed: %
\begin{align}
  s_{x t \nu} \coloneqq e^{\tau_{x t \nu}}\label{eq:priorsky} %
\end{align}
with $x \in \Omega_x$, $t \in \Omega_t$, and $\nu \in \Omega_\nu$ with $\mathcal P (\tau| T) \coloneqq \mathcal N (\tau\vert 0, T)$. 
The a priori correlation structure of the logarithmic sky brightness $\tau$ is encoded within the covariance $T$. %
Choosing a log-normal model allows the sky brightness to vary exponentially on linear spatial, temporal, and frequency scales and ensures the positivity of the reconstructed intensity, similarly to \cite{chael2016high, chael2018interferometric}. %

We perform a basis transformation to a standardised Gaussian distribution $\mathcal P(\xi_s) = \mathcal N(\xi_s\vert 0, \mathds{1})$, which allows us to separate the correlation structure from its realization \cite{repara}. %
The new coordinates $\xi_s$ have the same dimension as the original parameters, but are a priori independent: %
\begin{align}
  s = e^{A \xi_s} \quad \text{with} \quad A A^\dagger \coloneqq T .
\end{align}
This defines a generative model which turns standard normal distributed DOFs $\xi_s$ into random variables $s$ that are distributed according to \cref{eq:priorsky}. %
Although the information encoded in a distribution is invariant under coordinate transformations, MGVI depends on the choice of coordinates. %
Therefore, reformulating the entire inference problem in terms of standardised generative models is important to ensure that the prior information is fully captured by an approximation via MGVI. %
We visualize our generative model in \cref{fig:hierarchicalmodel}.

\subsection*{Correlations in space, time, and frequency}
We do not know the correlation structure of the logarithmic sky brightness a priori, so we include it as part of the model, which has to be inferred from the data. %
The different dimensions of the sky brightness are governed by completely distinct physical phenomena, which should be reflected in the model. %

Setting up such correlations involves a number of intricate technicalities. %
The main idea is to model the correlations in space, time, and frequency independently using the same underlying model and combine them via outer products. %
Doing this naively results in degenerate and highly un-intuitive model parameters. %
The model we introduce in the following avoids these issues, but unfortunately requires a certain complexity. %

For now we consider the correlation structure along the different sub-domains individually. %
A priori we do not want to single out any specific location or direction for the logarithmic sky brightness, which corresponds to statistical homogeneity and isotropy. %
According to the Wiener-Khinchin theorem, such correlation structures $T^{(i)}$ with $i\in\{\Omega_x,\Omega_t,\Omega_\nu\}$ are diagonal in the Fourier domain and can be expressed in terms of a power spectrum $p_{T^{(i)}}(|k|)$: %
\begin{equation}
	\begin{split}
		T^{(i)}_{k k'} &= \left(F^{(i)} T^{(i)}\left(F^{(i)}\right)^\dagger \right)_{k k'} \\&= \left(2 \pi\right)^{D^{(i)}} \delta\left(k - k'\right) \ p_{T^{(i)}}(|k|) , \\
	\end{split}
\end{equation}
where $F^{(i)}$ and $k$ denote the Fourier transformation and Fourier coordinates associated to the space $i$, $D^{(i)}$ is the dimension of $i$, $\delta$ denotes the Kronecker delta, and $|k|$ is the Euclidean norm of the vector $k$. %
We choose our Fourier convention such that no factors of $2 \pi$ enter the transformation $F^{(i)}$, and thus its inverse has a factor of $\nicefrac{1}{(2 \pi)^{D^{(i)}}}$. %
As we build the model in terms of standardised coordinates $\xi_s$, we work with the square root of the correlation matrix %
\begin{equation}\label{eq:amplitudeo}
	\begin{split}
		A^{(i)}_{k k'} &= \left(2 \pi\right)^{D^{(i)}} \delta\left(k - k'\right) \ \sqrt{ \ p_{T^{(i)}}(|k|)} \\&\eqqcolon \left(2 \pi\right)^D \delta\left(k - k'\right) p^{(i)}(|k|)
	\end{split}
\end{equation}
that converts those into the logarithmic brightness $\tau = A\,\xi_s$. %

The amplitude spectrum $p^{(i)}(|k|)$ depends on the characteristic length scales of the underlying physical processes, which we do not know precisely.  %
Our next task is to develop a flexible model for this spectrum that expresses our uncertainty and is compatible with a wide range of possible systems. %
We model the amplitude spectrum in terms of its logarithm: %
\begin{align}
 p^{(i)}(|k|)\propto e^{\gamma^{(i)}(|k|)}.
 \end{align}
We do not want to impose any functional basis for this logarithmic amplitude spectrum $\gamma^{(i)}(|k|)$, so we describe it non-parametrically using an integrated Wiener process in logarithmic $l = \text{log} |k|$ coordinates. %
This corresponds to a smooth, i.e.\ differentiable, function, with exponential scale dependence \cite{PhysRevD.83.105014}. %
In the logarithmic coordinates $l$, the zero-mode $|k|=0$ is infinitely far away from all other modes. Later on we deal with it separately and continue with all remaining modes for now. %

The integrated Wiener process in logarithmic coordinates $\gamma^{i}(l)$ reads:
\begin{align}
  \gamma^{(i)}(l) = m^{(i)} l + \eta^{(i)} \int_{l_0}^{l} \int_{l_0}^{l'} \xi^{(i)}_W(l'') \,dl' \,dl'' ,
\end{align}
where $l_0$ is the logarithm of the first mode greater than zero. %
Without loss of generality, we set the initial offset to zero. %
Later on we explicitly parameterise it in terms of a more intuitive quantity. %
The parameter $m^{(i)}$ is the slope of the amplitude on double-logarithmic scale. %
It is a highly influential quantity, as it controls the overall smoothness of the logarithmic sky brightness distribution. %
Specifically, after exponentiation, the spectrum is given as a power law with multiplicative deviations, and the exponent of this power law is given by the slope. %
Therefore, a spectrum with slope zero indicates the absence of any spatial correlation in the image, whereas a slope of $-1$ indicates continuous, and $-2$ differentiable brightness distributions along the respective axis \cite{oksendal2013stochastic}. %
The parameter $\eta^{(i)}$ describes how much the amplitude spectrum deviates from the power law. %
These deviations follow the smooth integrated Wiener process and can capture characteristic length scales of the logarithmic brightness distribution. %
Their precise shape is encoded in the realization $\xi_W^{(i)}\sim\mathcal{N}(\xi_W^{(i)}|0, \mathds{1})$, which are also parameters of our model and follow a priori the standard Gaussian distribution. %
We do not want to fix the slope and deviations and therefore impose Gaussian and log-normal priors for $j\in\{m, \eta\}$ respectively, with preference for a certain value $\mu_j^{(i)}$ and expected deviations $\sigma_j^{(i)}$ thereof: %
\begin{align}
m^{(i)} = \mu^{(i)}_{m} + \sigma^{(i)}_{m}\xi^{(i)}_{m}, \quad
\eta^{(i)}=e^{\mu^{(i)}_{\eta} + \sigma^{(i)}_{\eta}\xi^{(i)}_{\eta}}
\label{eq:hyperpara}
\end{align}
with $\xi^{(i)}_j\sim \mathcal{N}(\xi^{(i)}_j | 0,\mathds{1})$.

The amplitude spectrum defines the expected variation $\widetilde{U}^{(i)}$ of the log-brightness around its offset via
\begin{equation}
\widetilde{U}^{(i)} \coloneqq \int_{k \neq 0} p_{T^{(i)}}(|k|) \, dk = \int_{k \neq 0} e^{2\,\gamma^{(i)}(|k|)} \, dk .
\end{equation}
The relation between $\gamma^{(i)}$ and $\widetilde{U}^{(i)}$ is un-intuitive, but it is critical to constrain the expected variation to reasonable values as it has a severe impact on a priori plausible brightness distributions.
Therefore we replace the variance amplitude (i.e.\ the square root of $\widetilde{U}^{(i)}$) with a new parameter $a^{(i)}$: %
\begin{equation}
p^{(i)}(|k|) = a^{(i)}\, \frac{e^{\gamma^{(i)}(|k|)}}{\sqrt{\widetilde{U}^{(i)}}}, \quad \forall k \neq 0 .
\end{equation}
Note that this step implicitly determines the offset of the Wiener processes in terms of $a^{(i)}$. %
We elevate $a^{(i)}$ to be a free model parameter and impose a log-normal model analogous to $\eta^{(i)}$ with hyperparameters $\mu_a^{(i)}$ and $\sigma_a^{(i)}$. %

Next, we combine correlation structures in independent sub-domains. %
For every one of those, i.e.\ in our case space, time, and frequency, we use an instance of the model described above. %
We have not yet specified how to deal with the amplitude of the zero-modes $p^{(i)}(0)$, and their treatment emerges from the combination of the sub-domains. %
The overall correlation structure including all sub-domains is given by the outer product of the sub-spaces: %
\begin{align}
A = \bigotimes_{i\in\{x,t,\nu\}} A^{(i)}.
\end{align}
This product introduces a degeneracy: $\alpha (A^{(i)}\otimes A^{(j)})= (\alpha A^{(i)})\otimes A^{(j)}= A^{(i)}\otimes (\alpha A^{(j)})$ for all $\alpha\in\mathbb{R}^+$.
With every additional sub-domain we add one additional degenerate degree of freedom. %
We can use this freedom to constrain the zero-mode of the amplitude spectrum, and thus remove the degeneracy up to a global factor. %
For this we normalize the amplitudes in real-space: %
\begin{align}
	\begin{split}
		\widetilde{A}^{(i)}  &\coloneqq \left(\frac{1}{V^{(i)}} \int_{\Omega^{(i)}} \left(F^{(i)}\right)^{-1} p^{(i)} \ \text{d}\Omega^{(i)} \right)^{-1} A^{(i)} \\ &= \frac{V^{(i)}}{p^{(i)}(0)} A^{(i)} .
	\end{split}
\end{align}
The zero-mode of the normalised amplitude $\widetilde{A}^{(i)}$ can be fixed to the total volume $V^{(i)}$ of the space $\Omega^{(i)}$. %
Consequently, the overall correlation structure is expressed as %
\begin{align}\label{eq:spectra_amplitude}
A = \alpha \bigotimes_{i\in\{x,t,\nu\}} \widetilde{A}^{(i)} .
\end{align}
The remaining multiplicative factor $\alpha$ globally sets the scale in all sub-domains and has to be inferred from the data. %
Additionally, we put a log-normal prior with logarithmic mean $\mu_\alpha$ and standard deviation $\sigma_\alpha$ hyperparameters and a corresponding standard Gaussian parameter $\xi_\alpha$ on this quantity. %

This was the last ingredient for the correlation structure along multiple independent sub-domains and serves as a generative prior to infer the correlation structure in a space-time-frequency imaging problem. %
For the specific application to the EHT observations, however, only data averaged down to two narrow frequency channels is available. %
Therefore, as we do not expect to be able to infer a sensible frequency correlation structure using only two channels, we simplify \cref{eq:spectra_amplitude} to explicitly parameterize the frequency correlations as %
\begin{equation}
	A = \begin{pmatrix}
	1 & \epsilon \\
	1 & -\epsilon
	\end{pmatrix} \left(\alpha \bigotimes_{i\in\{x,t\}} \widetilde{A}^{(i)} \right) \ ,
\end{equation}
where $\epsilon$ is a hyperparameter that steers the a priori correlation between the frequency channels. %

We briefly summarise all the required hyperparameters and how the generative model for the correlation structure is built. %
We start with the correlations in the individual sub-domains which we describe in terms of their amplitude spectra $A^{(i)}(\xi^{(i)})$. %
Four distinct standardised model parameters are inferred from the data, $\xi^{(i)} \coloneqq (\xi^{(i)}_m ,\xi^{(i)}_\eta ,\xi^{(i)}_W, \xi^{(i)}_a )$. %
The first describes the slope of the linear contribution to the integrated Wiener process. %
The second is related to the strength of the smooth deviations from this linear part. %
The third parameter describes the actual form of these deviations. %
Finally, the last one describes the real-space fluctuations of the associated field. %

The hyperparameters are $\mu^{i}_j$ and $\sigma^{i}_j$  for $j \in \{ m,\eta, a\}$ specifying the expected mean and standard deviation of the slope $m^{(i)}$ and expected mean and standard deviation for $\ln (\eta), \ln (a)$, which are therefore enforced to be positive. %
In addition to these, we have to determine the global scale parameter $\alpha(\xi_\alpha)$, for which we also specify the logarithmic mean $\mu_\alpha$ and standard deviation $\sigma_\alpha$. %
We determine the values for the hyperparameters of the logarithmic quantities through an additional moment matching step by explicitly specifying the mean and standard deviation of the log-normal distribution.
%We set these hyperparameter means and standard deviations in case of the logarithmic quantities through a moment matching, specifying their mean and standard deviation on non-logarithmic scale instead.
The generative model for the correlation structure is therefore: %
\begin{align}
A(\xi_A) = \begin{pmatrix}
1 & \epsilon \\
1 & -\epsilon
\end{pmatrix} \left(\alpha(\xi_\alpha) \bigotimes_{i\in\{x,t\}} \widetilde{A}^{(i)}(\xi^{(i)})\right) 
\end{align}
with
\begin{align}
	\xi_A = \left(\xi_\alpha, \xi^{(x)}, \xi^{(t)}\right).
\end{align}
Combining this with the generative model for the sky brightness itself we end up with the full model: %
\begin{align}
s(\xi) = e^{F^{-1} \left(A(\xi_A)  \ \xi_s\right)}
\end{align}
with
\begin{align}
	F^{-1} = \bigotimes_{i\in\{x,t\}} \left(F^{(i)}\right)^{-1} \ . \label{eq:skymodel}
\end{align}
Our model is now standardized and all its parameters  $\xi = (\xi_A, \xi_s)$ follow a multivariate standard Gaussian distribution. %
The Bayesian inference problem is fully characterised by the negative logarithm (or information) of the joint probability distribution of data and parameters. %
Combining the closure likelihoods with the described sky brightness model therefore yields: %
\begin{multline}
		- \log\Big(\mathcal{P}(\varrho_d, \varphi_d , \xi)\Big) =\\
		=\frac12 \Big(\varrho_d -\varrho(s(\xi)) \Big)^\dagger
\Lambda_\rho^{-1}\Big(\varrho_d -\varrho(s(\xi)) \Big)\\
		+\frac12 \Big(\varphi_d-\varphi(s(\xi))\Big)^\dagger
\Lambda_\phi^{-1}\Big(\varphi_d-\varphi(s(\xi))\Big)\\
		+ \frac{1}{2}\xi^\dagger \xi+H_0 ,
\label{eq:ham}
\end{multline}
where $H_0$ is a constant that is independent of the latent variables $\xi$. %

\subsection*{Metric Gaussian Variational Inference}
So far, we have developed a probabilistic model in the generative form of the joint distribution of data and model parameters. %
In the end we want to know what the data tell us about the model parameters, as given in the posterior distribution according to Bayes' theorem. %
Our model is non-conjugate and we cannot solve for the result analytically. %
Instead, we approximate the true posterior distribution with a Gaussian using variational inference. %

This is fundamentally problematic, as we are approximating a multimodal posterior, which has multiple local optima, with a unimodal distribution. %
In the end, only one mode of the posterior will be captured by the variational distribution, underestimating the overall uncertainty. %
Some of these solutions can be considered equivalent.  %
For example, the absolute source location is neither constrained by the closure phases nor by the prior, but it is also irrelevant for the analysis. %
However, this shift-invariance also introduces several unphysical and pathological modes in the posterior, which might have low probability mass, but are local optima. %
An example for this is the appearance of multiple or partial copies of the source all over the image. %

Every reconstruction method that performs local optimization in the context of closure quantities potentially runs into these issues and our approach is no exception. %
Our chosen method and several procedures in our inference heuristic partially mitigate these issues and provide robust results. %
While we do not observe these pathological features in our main results, they do occur in the hyperparameter validation (see below).
One principled way to overcome them is posterior sampling, but the scale of the envisioned inference task with \num{7.4e6} parameters is prohibitively large.

We use Metric Gaussian Variational Inference (MGVI), which allows us to capture posterior correlations between all model parameters, despite the large scale of the inference problem.
MGVI is an iterative scheme that performs a number of subsequent Gaussian approximations $\mathcal{N}(\xi|\bar{\xi},\Xi)$ to the posterior distribution. %
Instead of inferring a parametrised covariance, an expression based on the Fisher information metric evaluated at the intermediate mean approximations is used, i.e.\ $\Xi \approx I(\xi)^{-1}$, with
\begin{multline}
		I(\xi) = \frac{\partial \varrho(s(\xi))}{\partial \xi}N_\varrho^{-1}\left(\frac{\partial \varrho(s(\xi))}{\partial \xi}\right)^\dagger \\ +
\frac{\partial e^{i\varphi(s(\xi))}}{\partial \xi}N_\varphi^{-1}\left(\frac{\partial e^{i\varphi(s(\xi))}}{\partial \xi}\right)^\dagger + \mathds{1} \ .
\end{multline}
The first two terms originate from the likelihood and the last from the prior. All of these are expressed in terms of computer routines and we do not have to store this matrix explicitly.
This is a non-diagonal matrix capturing correlations between all parameters.
To infer the mean parameter $\bar{\xi}$ we minimise the Kullback-Leibler divergence between the true posterior and our approximation:
\begin{multline}
\mathcal{D}_{\text{KL}}(\mathcal{N}(\xi|\bar{\xi},\Xi)||\mathcal{P}(\xi|\varphi_d,\varrho_d)) =\\= \int \text{d}\xi \ \mathcal{N}(\xi|\bar{\xi},\Xi) \ \text{ln}\left(\frac{\mathcal{N}(\xi|\bar{\xi},\Xi)}{\mathcal{P}(\xi \vert\varphi_d,\varrho_d )}\right).
\end{multline}
This quantity is an expectation value over the Gaussian approximation and measures the overlap between the true posterior and our approximation. %
As we minimise this quantity, the normalisation of the posterior distribution is irrelevant and we can work with the joint distribution over data and model parameters, as given by \cref{eq:ham}. %
We estimate the KL-divergence stochastically by replacing the expectation value through a set of samples from the approximation. %
The structure of the implicit covariance approximation allows us to draw independent samples from the Gaussian for a given location: %
\begin{align}
\xi^* \sim \mathcal{N}(\xi \vert 0, \Xi) \text{, therefore } \bar{\xi} \pm \xi^{*} \sim \mathcal{N}(\xi \vert \bar{\xi}, \Xi) .
\end{align}
Using the mean of the Gaussian plus and minus samples corresponds to antithetic sampling \cite{kroese2013handbook}, which reduces the sampling variance significantly, leading to performance increases. %
MGVI now alternates between drawing samples for a given mean parameter and optimising the mean given the set of samples. %
The main meta-parameters of this procedure are the number of samples and how accurately the intermediate approximations are performed. %

The procedure converges once the mean estimate $\bar{\xi}$ is self-consistent with the approximate covariance. %
To minimise the KL-divergence, we rely on efficient quasi-second-order Newton-Conjugate-Gradient in a natural gradient descent scheme. %
In the beginning of the procedure, the accuracy of KL and gradient estimates, as well as overall approximation fidelity, is not as important. %
In practice we gradually increase the accuracy with the number of MGVI iterations to gain overall speedups. %

\subsection*{Implementation Details}
We implement the generative model in NIFTy \cite{nifty5}, which also provides an implementation of MGVI utilising auto-differentiation.
We represent the spatial domain with $256\times 256$ pixels, each with a length of \SI{1}{\muas}. %
In the time domain we choose a resolution of $6$ hours for the entire observation period of $7$ days, thus obtaining $28$ time frames. %
The implementation of the generative model utilizes the Fast Fourier Transform and thus defines the resulting signal on a periodic domain. %
To avoid artefacts in the time domain, we add another $28$ frames to the end of the observed  interval, resulting in a temporal domain twice that size. %

For the frequency domain, only two channels are available, and we do not expect them to differ much from each other. %
Instead of inferring the correlation along this direction, as we do for the spatial and temporal axis, we assume a correlation between the two channels on the \SI{99}{\percent} level a priori, i.e.\ we set $\epsilon = 0.01$. %

This adds another factor of $2$ of required pixels to the reconstruction. %
For future reconstructions with deeper frequency sampling we can extend the model and treat this domain equivalently to the space and time domains. %
Overall we have to constrain $256\times256\times56\times2 + \text{power spectrum DOFs} \approx \num{7.4e6}$ pixel values with the data. %

The Gaussian approximation to the closure likelihoods is only valid in high signal-to-noise regimes \cite{closure19}. %
We increase the signal-to-noise ratio by means of an averaging procedure, which subdivides each individual scan into equally sized bins with a length of approximately 2 min.
To validate that this averaging is justified we compare the empirical standard deviation of averaged data values with the corresponding thermal noise standard deviation and find their ratio to be $\chitwovalue$ on average, consistent with the expected $\sqrt{2}$ for complex valued data.

The intra-site baselines of ALMA--APEX and SMT--JCMT probe the sky at scales larger than our field of view.
To avoid contamination from external sources, we flag these intra-site baselines and exclude closure quantities that involve the respective pair.

\subsection*{Hyperparameters}
The hyperparameter choices for the presented reconstruction are given in \cref{table:hyperparameters}. %
All hyperparameters except $\epsilon$ come in pairs of mean $\mu$ and standard deviation $\sigma$, parametrizing a Gaussian or log-normal distribution for a parameter.
This indirect hyperparameter setting induces a form of parameter search on each parameter, restricting them to be within a few standard deviations of the mean.

An exception to this is the frequency domain for which we only have two channels available.
Here, we set an a priori difference $\epsilon$ of \SI{1}{\percent}.
This is on the same order of magnitude as the relative difference in frequency, which is $\SI{0.9}{\percent}$.
The posterior can differ from this value, governed by the overall scale $\alpha$. %
This parameter controls the a-priori expected variance of the average logarithmic sky brightness mean and difference of the two frequencies.
For this overall scale $\alpha$, we set the mean $0.2$ with standard deviation $0.1$.
Since we normalize the flux of the final model, this parameter only controls the expected deviations of $\epsilon$, and has no other major effect.
A deviation of about half an $e$-fold would be expected with these hyperparameter settings, as it corresponds to the sum of two means.

Our choices regarding the remaining hyperparameter setting are motivated by being maximally agnostic with respect to the magnitude and shape of spatial correlations, while fixing the temporal correlations to be moderate.
By constraining the a priori slope of the spatial amplitude to $\mu_m^{x}=-1.5$ with a standard deviation of $\sigma_m^{(x)}=0.5$ we allow the model to express structures ranging from the rough Wiener process to the smooth integrated Wiener process within one standard deviation. %
The overall variance of the logarithmic sky brightness with respect to its spatial mean is set to be a-priori log-normal distributed with mean $0.7$ and standard deviation $1$.
A standard deviation larger than its mean induces a log-normal distribution with a heavy tail, thus allowing for potentially large posterior spatial fluctuations.

The flexibility parameter $\eta$ specifies the degree to which the power spectrum can deviate from a power-law shape and thereby introduce characteristic length- or time-scales.
We choose small values for its mean ($0.01$) and standard deviation ($0.001$), discouraging such characteristic scales in both time and space. %
Still if necessary, strong deviations from a power law are possible if the data demand it (see \cref{fig:ps}).

In the time domain we do not expect strong variability due to the physical scale of the system, extending over several light-days. %
We express this through the slope of the temporal amplitude, setting its expected mean to $\mu_m^{(t)}=-4$ and standard deviation $\sigma_m^{(t)}=0.5$, imposing long correlations in time.
The overall fluctuations are again relatively unconstrained with mean $0.2$ and standard deviation $1$. %

To test the sensitivity of our method, we perform a dedicated hyperparameter study in a later paragraph.

% Even though we have this parameter restricted, we still match the validation
% examples quite well

\subsection*{Inference Heuristic}
Here we want to give the motivation behind the choices for our inference heuristic, as it is described in \cref{tab:implementation}. %
These are ad-hoc, but using the described procedure provides robust results for all examples given the described set of hyperparameters. %

Our initial parametrization corresponds to a signal configuration that is constant in time and shows a Gaussian shape centred in the field of view with standard deviation of \SI{30}{\muas}. %
This breaks the translation symmetry of the posterior distribution, concentrating the flux towards the centre. %
It does not fully prevent the appearance of multiple source copies, but they are not scattered throughout the entire image. %
A similar trick is also employed in the EHT-Imaging pipeline. %

The next issue we are facing is \enquote{source teleportation}. %
Close-by frames are well-constrained by our assumed correlation, but the data gap of four days allows for solutions in which the source disappears at one place and re-appears at another. %
This is also due to the lack of absolute position information and not prevented by our dynamics prior. %
To avoid these solutions, we start by initially only using data of the first two days. %
For these we recover one coherent source, which is extrapolated in time. %
Once we include the data of the remaining two days, the absolute location is already fixed and only deviations and additional information to previous times have to be recovered. %

The appearance of multiple source copies can be attributed to multi-modality of the posterior. %
The stochastic nature of MGVI helps, to some degree, to escape these modes towards more plausible solutions. %
Nevertheless, this is not enough for strongly separated optima. We therefore employ a tempering scheme during the inference. %
The phases constrain the relative locations in the image, whereas the amplitudes constrain the relative brightness. %
Smoothly aligning source copies while keeping the amplitudes constant is either impossible or numerically stiff. %
Allowing to violate the observed closure amplitudes for a short period of time makes it easier to align all copies to a single instance. We achieve this by not considering the closure amplitude likelihood during one intermediate step of MGVI. %
The same issue persists for the closure amplitudes. %
We therefore alternate between only phase-likelihood and amplitude-likelihood. %
In between these two we always perform a step using both likelihoods.
We start this procedure after a fixed number of steps, allowing a rough source shape to form beforehand. %
In the end we use the full likelihood for several steps.

MGVI requires specifying the number of sample pairs used to approximate the KL-divergence. %
The more samples we use, the more accurate the estimate, but the larger the overall computational load. %
We steadily increase the number of samples throughout the inference for two reasons.
Initially the covariance estimate only inaccurately describes the posterior mode and a large number of samples would be a waste of computational resources.
Additionally, fewer samples increase the stochasticity of the inference, which makes it more likely to escape pathological modes of the posterior. %
Towards the end, it is worth investing computational power into a large number of samples in order to obtain accurate uncertainty estimates.

Finally, we have to specify how and how well the KL is optimized in every MGVI step. %
In the beginning, we do not want to optimize too aggressively, as we only use a limited number of samples and we want to avoid an over-fitting on the sample realizations. %
We therefore use the LBFGS \cite{lbfgs} method with an increasing number of steps. %
For the last period, where we have accurate KL estimates, we employ the more aggressive natural gradient descent equivalent to \texttt{scipy}'s \texttt{NewtonCG} algorithm \cite{newtoncg} to achieve deep convergence.

To demonstrate the robustness of this procedure we perform the reconstruction of M87* and the six validation examples (see below) for five different random seeds, in total 35~full reconstructions.
Using the described heuristic, we do not encounter any of the discussed pitfalls, and we obtain consistent results.
This corresponds to a success rate of at least \SI{97}{\percent}.

\begin{table*}
  \centering
  \begin{tabular}{l|rr|rr}
    \toprule %
      Parameter & mean & std.\ deviation & log-mean & log-std.\ deviation \\\midrule
      $\alpha$ & 0.2 & 0.1 & $\mu_{\alpha}^{(x)}=-1.7$ & $\sigma_{\alpha}^{(x)}=0.47$\\ %
      $a$ & 1.5 & 1. & $\mu_{a}^{(x)}=0.22$ & $\sigma_{a}^{(x)}=0.61$\\ %
      $m$ & $\mu_{m}^{(x)}=-1.5$ & $\sigma_{m}^{(x)}=0.5$ & N/A & N/A\\
      $\eta$ & 0.01 & 0.001 & $\mu_{\eta}^{(x)}=-4.6$ & $\sigma_{\eta}^{(x)}=0.10$\\ %
      \midrule %
      $a$ & 0.2 & 1. & $\mu_{a}^{(t)}=-3.2$ & $\sigma_{a}^{(t)}=1.8$\\ %
      $m$ & $\mu_{m}^{(t)}=-4$ & $\sigma_{m}^{(t)}=0.5$ & N/A & N/A\\
      $\eta$ & 0.01 & 0.001 & $\mu_{\eta}^{(t)}=-4.61$ & $\sigma_{\eta}^{(t)}=0.10$\\ %
      \midrule %
      $\epsilon=0.01$ & N/A & N/A & N/A & N/A\\ % expected frequency deviation

    \bottomrule
  \end{tabular}
  \caption[Hyperparameters of the generative model]{Hyperparameters of the generative model.
The first column indicates the symbol of the parameter, as used in the manuscript.
The second and third column denote the a priori mean and standard deviation.
All quantities for which positivity is enforced are modelled as log-normal distribution.
Their corresponding logarithmic mean and logarithmic standard deviation are reported in the fourth and fifth columns.
    The expected frequency deviation $\epsilon$ is fixed and not variable.
    \label{table:hyperparameters}}
\end{table*}

\begin{table*}
	\usetikzlibrary{arrows,calc,decorations.markings,math,arrows.meta, decorations.pathmorphing,backgrounds,positioning,fit,matrix}

	\newcommand\tikzmark[2]{%
		\tikz[remember picture,baseline] \node[above, outer sep=0pt, inner sep=0pt] (#1){\phantom{#2}};%
	}
\newcommand\link[2]{%
	\begin{tikzpicture}[remember picture, overlay, >=Stealth, shift={(0,0)}]
	\draw[double, -Latex, line width=0.5mm] (#1) to (#2);
	\end{tikzpicture}%
}
\centering

\begin{tabular}{|c|r|r|r|r|}
  \toprule
  Iteration & Data Set& Tempering & Optimizer & Sample Pairs \\\midrule
  $i=0$&$i \geq 0$&$i \geq 0$&$i \geq 0$&$i \geq 0$ \\
  \tikzmark{a}{$i=1$} &&  & &  \\
    && full likelihood & & 			 \\
  &first two days&   &  &  \\\cline{3-3}
   && $i\geq10$ &V-LBFGS  & 			 \\
   && & $4*(4+i//4)$& 			 \\
   &&  &iterations & 			 \\\cline{2-2}
   &$i\geq30$& alternating & & 			 \\
   & &  & & 	$N=10 * (1+i//8)$	\\
  &&  & & 			 \\
  &&  & & 			 \\
   &&  & & 			 \\\cline{3-4}
  &all days& $i\geq50$ & $i\geq50$& 			 \\
   &&  & & 			 \\
     &&  & Natural Gradient& 			 \\
       && full likelihood &$20$ iterations & 			 \\
  \tikzmark{b}{$i=1$}  &&  & & 			 \\
   $i=59$&&  & & 			 \\
  \bottomrule
\end{tabular}
\link{a}{b}

\caption[Minimisation scheme used for the inference]{Minimisation scheme used for the inference. In addition to the mentioned samples, their antithetic counterparts were used as well.}
\label{tab:implementation}
\end{table*}

\subsection*{Method validation}
\paragraph{Synthetic observations}
We validate our method on six synthetic examples, three of which exhibit temporal variation. %
The first two time-variable examples are crescents with an evolution of the angular asymmetry on time scales similar to what was measured by the EHT collaboration for M87*.
They are toy models of the vicinity of the black hole and are defined analogously to \cite[Section~C.2]{ehtiv}:
\begin{multline}
b_0(r_0,A, w;x, y, t) \propto \exp\left(-\frac{(\sqrt{x^2+y^2} - r_0)^2}{2\,(w/2.355)^2} \right)\cdot\\\cdot \left( 1+2 A \sin\left[ \arctan\left(\frac{y}{x}\right) + \SI{240}{\degree} + \frac{\SI{20}{\degree}}{\SI{7}{\day}}\, t \right] \right),
\end{multline}
where $r_0$ is the ring radius, $A$ the ring asymmetry, $w$ the full width half maximum of the ring, and $x$, $y$, and $t$ are space and time coordinates.
We choose two sets of parameters.
The first, called \texttt{eht-crescent}, follows the validation analysis of the EHT Collaboration \cite{ehtiv}: $r_0 = \SI{22}{\muas}$, $A=0.23$, and $w= \SI{10}{\muas}$.
The second, called \texttt{slim-crescent}, has a smaller radius, a more pronounced asymmetry, and a sharper ring: $r_0 = \SI{20}{\muas}$, $A = 0.5$, and $w=\SI{3}{\muas}$.

As a third example, called \texttt{double-sources}, we choose two Gaussian shapes $b(t, x, y)$ with full-width half maximum $r=\SI{20}{\muas}$ that approach each other:
\begin{align}
\tilde b_1 (x_0, y_0; t, x, y) &= \exp \left( -\frac{(x-x_0)^2+(y-y_0)^2}{2\,(r/2.355)^2} \right),
\end{align}
\begin{multline}
b_1(t,x, y) \propto \tilde b_1 (\alpha \sin (\phi), \alpha \cos (\phi); t, x, y) +\\+ \tilde b_1 (-\alpha \sin (\phi), -\alpha \cos (\phi); t, x, y),
\end{multline}
where $x$, $y$, and $t$ are space and time coordinates, $\alpha$ is the time-dependent distance, and $\phi$ the time-dependent angle:
\begin{align}
	\alpha (t) &= \SI{32}{\muas} - \frac{\SI{6}{\muas}}{\SI{7}{\day}} \,t,\\
	\phi (t) &=  \frac{\pi}{12}\frac{t}{\SI{7}{\day}} -\SI{21.8}{\degree}.
\end{align}
The static examples consist of a uniform disk with blurred edges and two simulations of black holes, \texttt{challenge1} and \texttt{challenge2}, taken from the EHT imaging challenge \cite{vlbiimagingchallenge}.
The brightness of the blurred \texttt{disk} with a diameter of \SI{40}{\muas} is given by:
\begin{align}
  b_2(x,y) \propto \frac{1}{2}\left(1+\tanh \left[\frac{\SI{20}{\muas}-\sqrt{x^2+y^2})}{\SI{3}{\muas}}\right]\right),
\end{align}
where $x$ and $y$ again denote the spatial coordinates.

For our validation we simulate the M87* observation, using the identical uv-coverage, frequencies, and time sampling. %
We set the total flux of the example sources to \SI{0.5}{\jy} and add the reported thermal noise from the original observation. %
We do not add non-closing errors, such as polarization leakage.
We also ignore the existence of large-scale emission around the source, as it would be expected for M87* \cite{kim2018limb}.
This kind of emission only has a significant contribution to the intra-site baselines \cite{ehtiv}.
By excluding these, we make sure that the large-scale emission does not affect our results.
The reconstruction follows the identical procedure as for M87*, using the same hyperparameters and pixel resolution. %

The results of the dynamic examples versus the ground truth and the pixel-wise uncertainty are shown in \cref{fig:validation}. %
For all static examples, we do not find time-variability in the reconstructions.
Thus, we only show the first frame versus ground truth, smoothed ground truth, and the pixel-wise uncertainty in the figure. %
As the likelihood is invariant under shifts, offsets in the reconstruction are to be expected. %
We are able to recover the shapes of the different examples, irrespective of the source being static or not. %

\begin{table*}
  \centering
	
\begin{tabular}{lccccc}
\toprule
& $d\, (\mu\text{as})$ & $w\, (\mu\text{as})$ & $\eta\, (^\circ )$ & $A$ & $f_C$\\\midrule
\multicolumn{6}{l}{\textsc{Ground truth (uncertainty as per \cite[Table 7]{ehtiv})}}\\
April 5 & $44.5 \pm 0.7$ & $10.0 \pm 0.8$ & $150.0 \pm 0.0$ & $0.23 \pm 0.00$ & $0.000 $ \\
April 6 & $44.5 \pm 0.7$ & $10.0 \pm 0.8$ & $152.9 \pm 0.0$ & $0.23 \pm 0.00$ & $0.000 $ \\
April 10 & $44.5 \pm 0.7$ & $10.0 \pm 0.8$ & $164.3 \pm 0.0$ & $0.23 \pm 0.00$ & $0.000 $ \\
April 11 & $44.5 \pm 0.7$ & $10.0 \pm 0.9$ & $167.1 \pm 0.0$ & $0.23 \pm 0.00$ & $0.000 $ \\\midrule
\multicolumn{6}{l}{\textsc{Our method (uncertainty as per \cite[Table 7]{ehtiv})}}\\
April 5 & $43.8 \pm 2.8$ & $16.4 \pm 3.6$ & $152.4 \pm 3.4$ & $0.24 \pm 0.04$ & $0.192 $ \\
April 6 & $43.8 \pm 2.7$ & $16.4 \pm 3.5$ & $150.9 \pm 6.5$ & $0.23 \pm 0.05$ & $0.193 $ \\
April 10 & $43.9 \pm 2.7$ & $16.5 \pm 3.8$ & $165.5 \pm 1.1$ & $0.22 \pm 0.05$ & $0.193 $ \\
April 11 & $43.9 \pm 2.7$ & $16.6 \pm 4.0$ & $168.6 \pm 2.1$ & $0.22 \pm 0.05$ & $0.193 $ \\\midrule
\multicolumn{6}{l}{\textsc{Our method (sample uncertainty)}}\\
April 5 & $43.5 \pm 0.8$ & $15.6 \pm 1.9$ & $152.2 \pm 4.7$ & $0.23 \pm 0.02$ & $0.196 \pm 0.052$\\
April 6 & $43.5 \pm 0.8$ & $15.6 \pm 1.9$ & $152.9 \pm 4.4$ & $0.23 \pm 0.02$ & $0.195 \pm 0.052$\\
April 10 & $43.6 \pm 0.8$ & $15.7 \pm 1.9$ & $166.6 \pm 4.6$ & $0.23 \pm 0.02$ & $0.196 \pm 0.052$\\
April 11 & $43.6 \pm 0.8$ & $15.8 \pm 1.9$ & $169.2 \pm 4.8$ & $0.23 \pm 0.02$ & $0.196 \pm 0.053$\\\bottomrule
\end{tabular}

	\caption[Crescent parameters recovered for the \texttt{eht-crescent} example]{
    The crescent parameters recovered from the \texttt{eht-crescent} validation example versus ground truth.
    Analogue to \cref{tab:ringfits}.
  }
	 \label{tab:ehtcrescentvalidation}
\end{table*}

\begin{table*}
\centering

\begin{tabular}{lccccc}
\toprule
& $d\, (\mu\text{as})$ & $w\, (\mu\text{as})$ & $\eta\, (^\circ )$ & $A$ & $f_C$\\\midrule
\multicolumn{6}{l}{\textsc{Ground truth (uncertainty as per \cite[Table 7]{ehtiv})}}\\
April 5 & $40.0 \pm 1.1$ & $7.0 \pm 1.4$ & $150.0 \pm 0.0$ & $0.50 \pm 0.00$ & $9.7 \times 10^{-7} $ \\
April 6 & $40.0 \pm 1.0$ & $7.0 \pm 1.3$ & $152.9 \pm 0.0$ & $0.50 \pm 0.00$ & $9.6 \times 10^{-7} $ \\
April 10 & $40.1 \pm 1.0$ & $7.1 \pm 1.3$ & $164.3 \pm 0.0$ & $0.50 \pm 0.00$ & $9.6 \times 10^{-7} $ \\
April 11 & $40.1 \pm 1.1$ & $7.2 \pm 1.4$ & $167.1 \pm 0.0$ & $0.50 \pm 0.00$ & $9.6 \times 10^{-7} $ \\\midrule
\multicolumn{6}{l}{\textsc{Our method (uncertainty as per \cite[Table 7]{ehtiv})}}\\
April 5 & $36.9 \pm 12.5$ & $18.8 \pm 10.7$ & $149.7 \pm 3.9$ & $0.44 \pm 0.08$ & $0.163 $ \\
April 6 & $37.1 \pm 12.3$ & $18.7 \pm 10.5$ & $151.3 \pm 2.6$ & $0.44 \pm 0.07$ & $0.160 $ \\
April 10 & $38.2 \pm 11.7$ & $18.7 \pm 9.9$ & $162.6 \pm 5.1$ & $0.45 \pm 0.09$ & $0.151 $ \\
April 11 & $38.3 \pm 11.3$ & $18.4 \pm 9.3$ & $163.5 \pm 5.5$ & $0.45 \pm 0.10$ & $0.154 $ \\\midrule
\multicolumn{6}{l}{\textsc{Our method (sample uncertainty)}}\\
April 5 & $37.2 \pm 1.1$ & $16.3 \pm 2.0$ & $149.7 \pm 3.7$ & $0.45 \pm 0.03$ & $0.181 \pm 0.051$\\
April 6 & $37.3 \pm 1.1$ & $16.2 \pm 2.0$ & $151.4 \pm 3.7$ & $0.45 \pm 0.03$ & $0.179 \pm 0.051$\\
April 10 & $37.9 \pm 1.2$ & $16.4 \pm 2.0$ & $164.0 \pm 4.0$ & $0.44 \pm 0.04$ & $0.175 \pm 0.049$\\
April 11 & $38.0 \pm 1.1$ & $16.5 \pm 2.0$ & $165.6 \pm 4.1$ & $0.44 \pm 0.04$ & $0.176 \pm 0.049$\\\bottomrule
\end{tabular}

\caption[Crescent parameters recovered for the \texttt{slim-crescent} example]{
  The crescent parameters recovered from the \texttt{slim-crescent} validation example versus ground truth.
    Analogue to \cref{tab:ringfits}.
  }
\label{tab:crescentvalidation}
\end{table*}

The ring-parameter analysis is applied to the two crescent scenarios as well.
The results for the recovered diameter $d$, width $w$ and orientation angle $\eta$ are shown in \cref{tab:ehtcrescentvalidation,tab:crescentvalidation}. %
Here we compare the ground truth to the analysis of the mean reconstruction, following the approach of the EHT collaboration. %
In order to propagate the uncertainty estimate of our reconstruction directly, we can extract the crescent parameters of all samples individually to obtain a mean estimate with associated uncertainty. %
The variational approximation has the tendency to under-estimate the true variance and in this case should be regarded more as a lower limit. %
For the estimation of the ring diameter we adopt the approach described in Appendix G of \cite{ehtiv} to correct the diameter for the bias due to finite resolution. %
We further discuss the recovered spatial correlation structures of the log-brightness in \cref{sec:powersp}.

\begin{figure*}
  \centering
  \includegraphics[width=\textwidth]{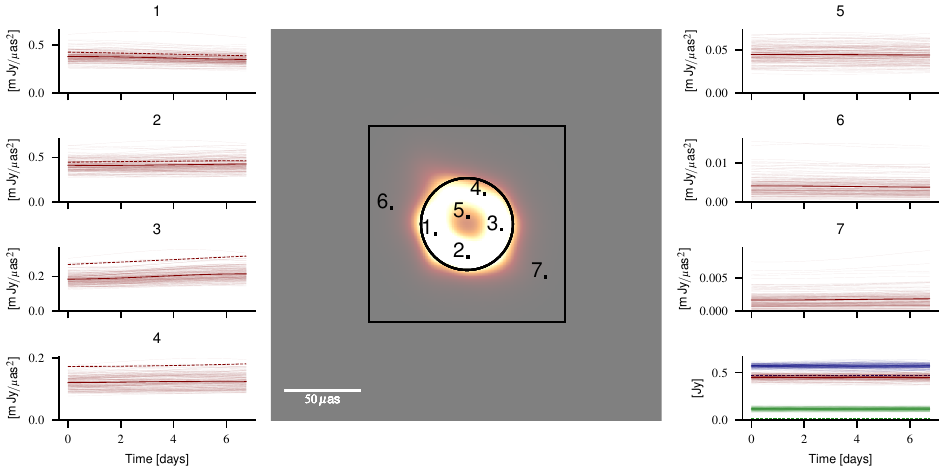}
  \caption[Time evolution of the validation data set \texttt{eht-crescent}]{The time evolution of the validation data set \texttt{eht-crescent}.
    Analogous to \cref{fig:timeseries}.
    The dashed lines represent the ground truth.
    In subfigures 5 to 7 the ground truth is constantly zero.
  }
  \label{fig:ehtcrescenttimeseries}
\end{figure*}

Starting with the first crescent, we recover well the diameter $d$, orientation angle $\eta$, and asymmetry $A$. %
The ground truth is within the uncertainty of both procedures.  %
The width $w$ of the crescent is below the angular resolution of the telescope, so it is not surprising that we do not fully resolve it in the reconstruction. Both ways to calculate the uncertainty do not account for the discrepancies.
Interestingly, all quantities, except for the orientation angle, are static in time.
For this example, we additionally show the temporal evolution of selected points in \cref{fig:ehtcrescenttimeseries}, analogously to M87*.
The reconstruction follows the dynamics of the ground truth, as indicated by the dashed line.

The reconstruction of the \texttt{slim-crescent} proves more challenging. %
Due to the weak signal, we do not recover the faint part of the circle. %
For an accurate extraction of the ring parameters, however, this area is vital to constrain the radius. %
Here, we only recover the orientation angle well. %
The diameter estimate has large error bars when following the approach of the EHT collaboration. %
In this scenario the uncertainty estimate appears too conservative. %
In contrast to that, using samples for the uncertainty provides significantly smaller error bars. %
This could be due to the variational approximation, which tends to under-estimate the true uncertainty. %

The dynamics of the two Gaussian shapes are recovered accurately and our model correctly interpolates through the gap of three days without data.

Overall, our method is capable of accurately resolving dynamics that are comparable to the ones expected in M87*.
Therefore, our findings regarding the temporal evolution of M87* may be trusted.

\Cref{fig:spectralindex} shows the relative spectral index of M87*, as well as the \texttt{eht-crescent} validation example. %
In both cases, an increased spectral index coincides with the brightest spot on the ring. %
This is not a feature of the validation example, as we use a constant spectral index throughout the source. %
The apparent feature could emerge from different coverage, as well as a bias due to the unimodal approximation. %
Nevertheless, these features are insignificant as our reported posterior uncertainty is large enough to be consistent with a constant spectral index throughout the image. %
This finding is not surprising due to the small separation of the two channels. %
In principle our method is capable of providing a spectral index, but in this application the data is inconclusive. %

The reconstructions of the three static examples are shown in \cref{fig:staticvalidation}.
For illustrative purposes we show not only the ground truth, but also a blurred image of the ground truth, which we obtain by convolving with a Gaussian beam of \SI{12}{\muas}.  % sigma = 5 px = 5 muas -> FWHM = 11.775
Overall we recover the general shape and main features of the sources. %

None of the validation reconstructions yield imaging artefacts that appear in any way similar to the elongated structure that our algorithm recovers in the south-western and north-eastern directions of M87*. %
%None of the validation reconstructions suffer from imaging artefacts similar to the elongated structures in the south-western and north-eastern direction of M87*. %
Especially the \texttt{eht-crescent} model is accurately recovered without a trace of spurious structures.
We conclude that the elongated features of M87* are either of physical origin or due to baseline-based errors and that they are not an artefact introduced by our imaging technique.

\paragraph{Hyperparameter validation}
To study the sensitivity of our results with regard to hyperparameters, we repeat the reconstruction of M87*, as well as \texttt{eht-crescent}, with 100 randomized, but shared configurations.
We do not vary the standard-deviation related hyperparameters, but sample the corresponding mean hyperparameters uniformly within three respective standard deviations.
For the expected frequency deviation $\epsilon$ we sample logarithmically uniformly between $0.001$ and $0.5$.
Some of these configurations are numerically unstable and will result in errors.
Other configurations do not facilitate the emergence of a single source and exhibit typical VLBI artefacts, especially source doubling throughout the image.
This behaviour is easy to detect and we label the results manually.

\begin{figure*}
    \centering
    \includegraphics{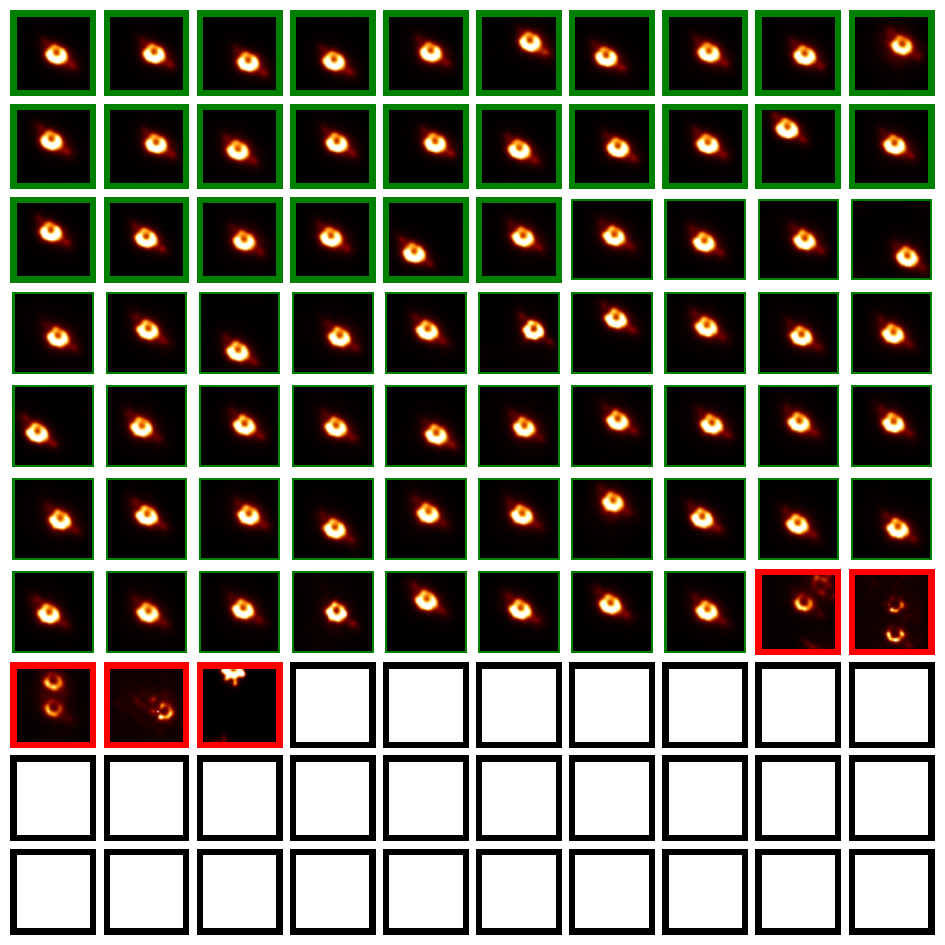}
    \caption[Results of the hyperparameter validation for M87*]{
        Approximate posterior mean of M87* for 100 different hyperparameter settings within $\pm3$ standard deviations around our chosen hyper parameters.
        The colored borders indicate the quality of the result: green for results that are visually similar to the main run, red for unphysical results, and black (with plain white image content) for runs in which the algorithm numerically diverged.
        The green thick and thin border corresponds to results with total reduced $\chi^{2}$ value below and above $1.2$, respectively.
    }
    \label{fig:allvalidationsm87}
\end{figure*}
  
\begin{figure*}
  \centering
  \includegraphics{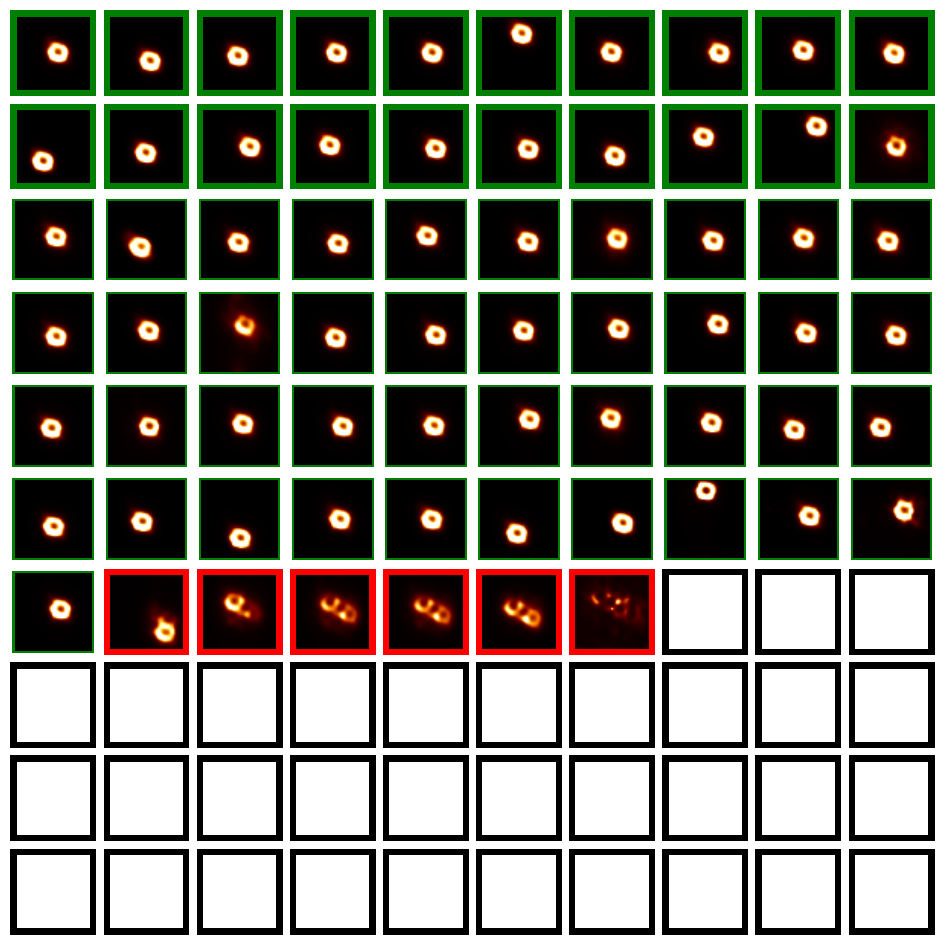}
  \caption[Results of the hyperparameter validation for \texttt{eht-crescent}]{
    Approximate posterior mean of the \texttt{eht-crescent} validation for 100 different hyperparameter settings, analogous to Supplementary \cref{fig:allvalidationsm87}.
  }
  \label{fig:allvalidationsehtcrescent}
\end{figure*}

The resulting mean sky brightness distributions can be found in \cref{fig:allvalidationsm87,fig:allvalidationsehtcrescent}.
The algorithm fails in 30\% of the cases, results in artefacts 5.5\% of the time, and facilitates the emergence of a single ring in 64.5\% of all cases.
All of the latter cases exhibit extended structures in the case of M87*, whereas we do not observe any similar features for \texttt{eht-crescent}.
We are therefore confident that these do not originate from the choice of the hyperparameters.

\begin{figure*}
  \centering
  \includegraphics{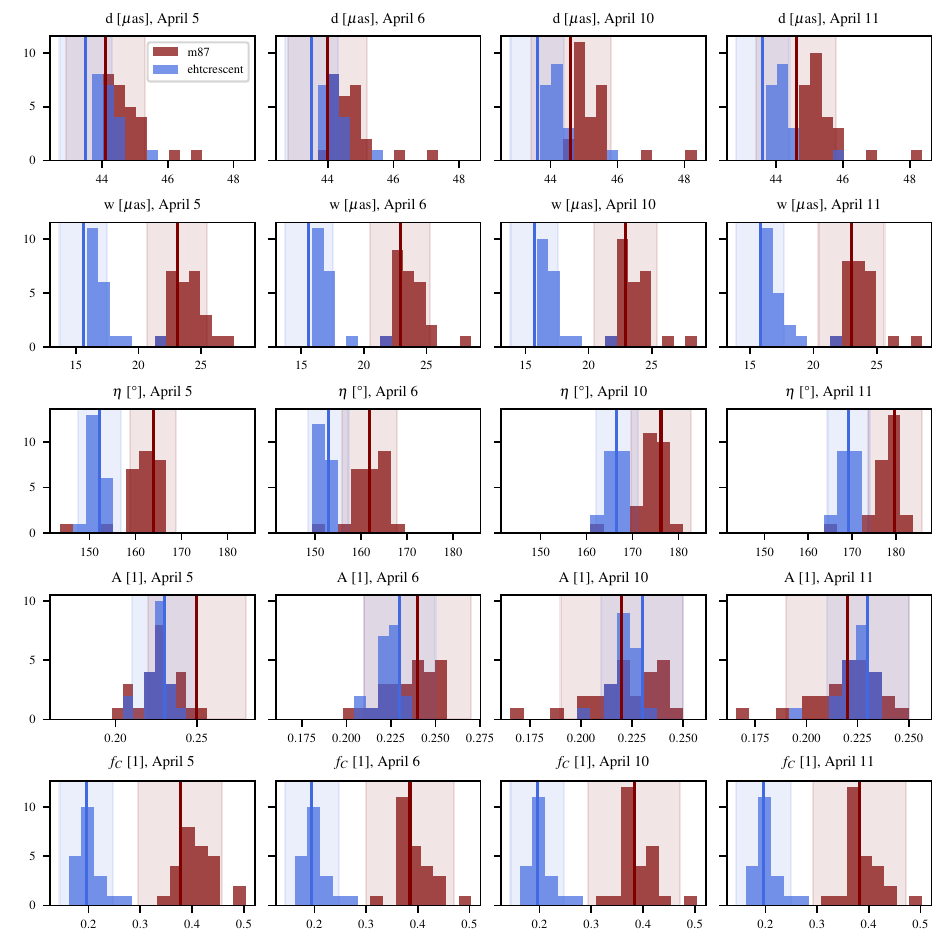}
  \caption[Crescent parameters from the hyperparameter validation]{
    Histograms of the crescent parameters over all hyper parameter validation runs with a total reduced $\chi^{2}$ value below $1.2$.
    The vertical lines and shaded area display the approximate posterior means and 1-$\sigma$ standard deviations as reported in \cref{tab:ringfits} and Supplementary \cref{tab:ehtcrescentvalidation}.
  }
  \label{fig:hyperparameterhistograms}
\end{figure*}

For a significant portion of the parameter configurations we do not find the shift of the brightness asymmetry.
However, the results with a static source all exhibit a significantly higher reduced $\chi^2$ value compared to the reconstructions that feature a shift in brightness asymmetry.
In 23\% of all test cases we obtain reconstructions with shifting asymmetry, all of which are consistent with the main result of this paper.
\Cref{fig:hyperparameterhistograms} shows that all reported ring fit parameters of our main result including their uncertainties are fully consistent with the hyperparameter validation.

There are two possible explanations for the absence of asymmetry shifts.
First, the prior on the temporal evolution already favours slow dynamics and sampling even more extreme values for this validation might lead to static reconstructions.
Second, the inference heuristic was optimized for the parameter sets similar to the one used for the main result and not for the large variety of cases.
They numerically pose completely different challenges and might converge more slowly or exhibit different optima.
Improvements in the heuristic would most probably lead to a more robust behaviour for a larger parameter range.

\paragraph{Data consistency}

\begin{figure*}
  \centering
  \includegraphics[width=\textwidth]{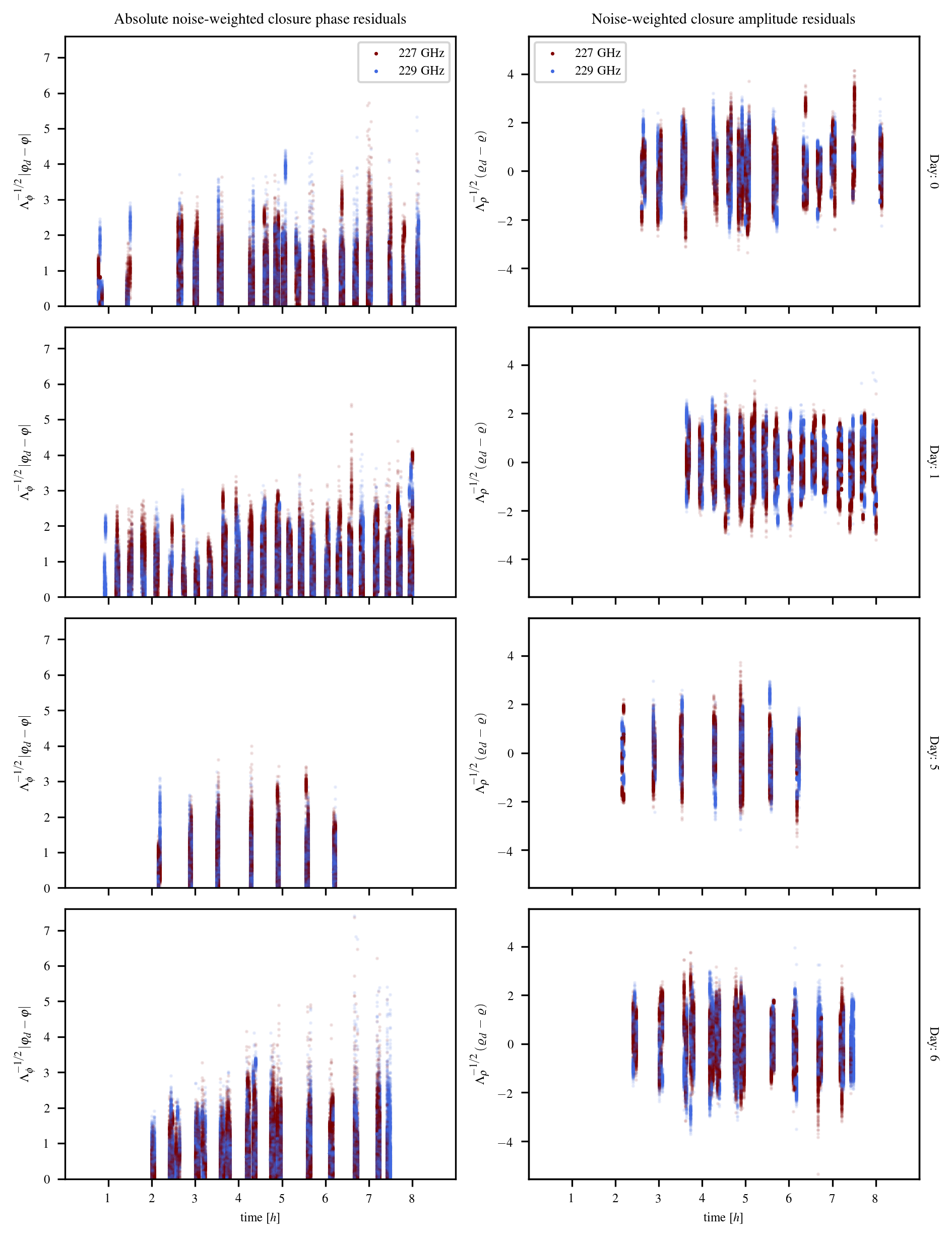}
  \caption[Absolute noise-weighted closure residuals for M87*]{
    Absolute noise-weighted residuals for the closure phases $\varphi$ (left column) and closure amplitudes $\varrho$ (right column) for the two frequency channels (\SI{227}{\giga\hertz} in red, and \SI{229}{\giga\hertz} in blue), for the M87* reconstruction for all samples. Each row corresponds to one of the four observational days and the x-axis of each figure corresponds to the time (in hours) progressed on the corresponding day.
  }
  \label{fig:m87residuals}
\end{figure*}

\begin{figure*}
  \centering
  \includegraphics[width=\textwidth]{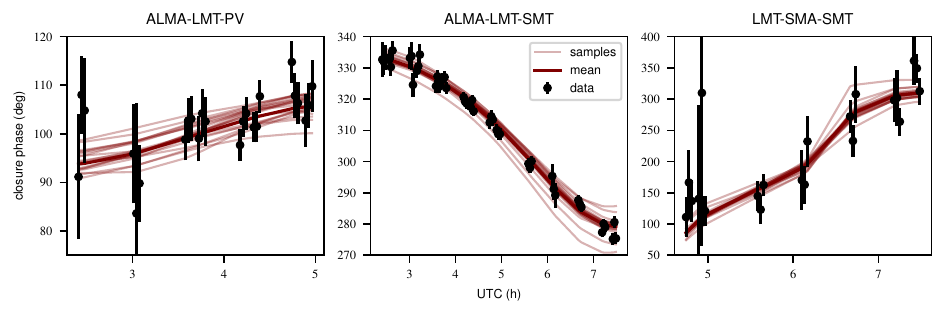}
  \caption[Closure phases for selected antenna triplets in time]{
    Closure phases for selected antenna triplets in time.
    The plot shows observations made at \SI{229}{\giga\hertz} (black dots) on April 11th with black lines indicating the 1-sigma measurement error, as well as artificial measurements of 20 uncurated posterior samples (transparent maroon lines) and the corresponding posterior mean (maroon line).
    Note that this data is not directly employed in our algorithm since we further process the data in order to take correlations between different closure phases into account, as described in the likelihood subsection of the methods section.
  }
  \label{fig:selectedresiduals}
\end{figure*}

\begin{table*}
  \centering
  \begin{tabular}{lrrrr}\toprule
& April 5 & April 6 & April 10 & April 11 \\\midrule
eht-crescent & $1.2, 1.0$ & $1.3, 0.9$ & $1.0, 0.9$ & $1.4, 1.1$\\
challenge1 & $1.2, 1.0$ & $1.3, 1.2$ & $1.4, 1.3$ & $1.1, 1.1$\\
challenge2 & $1.4, 0.9$ & $1.3, 0.9$ & $1.4, 0.9$ & $1.2, 0.9$\\
slim-crescent & $1.1, 1.1$ & $1.0, 1.0$ & $1.0, 1.0$ & $1.0, 1.0$\\
disk & $1.6, 1.2$ & $1.4, 1.3$ & $1.5, 1.4$ & $1.3, 1.2$\\
double-sources & $1.2, 1.1$ & $1.2, 1.1$ & $1.3, 1.3$ & $1.4, 1.1$\\
m87 & $1.1, 0.9$ & $1.1, 0.8$ & $1.1, 0.9$ & $1.1, 0.9$\\\midrule
m87 (EHT-imaging) & $1.0, 1.0$ & $1.0, 1.0$ & $1.0, 0.8$ & $1.0, 1.0$\\
\bottomrule
\end{tabular}

  \caption[Sample-averaged reduced $\chi^{2}$ values]{Sample-averaged reduced $\chi^{2}$ values. The left and right values are the reduced $\chi^{2}$ values for the closure phase and the closure amplitude likelihood, respectively.
    The reduced $\chi^{2}$ values for the EHT-imaging results are computed for the likelihood presented in this work.
    Note that the EHT-imaging results themselves have been obtained using a different likelihood and the reduced $\chi^{2}$ value is not a posterior average but computed using the mean image.
  }
  \label{tab:chisqtable}
\end{table*}

The time-resolved residuals-$\chi^2$ of the closure phases and amplitudes for all validation examples, as well as for M87* are shown in \cref{tab:chisqtable}. %
Additionally, in \cref{fig:m87residuals}, we display the noise-weighted residuals for the M87* reconstruction for the four observation periods as a function of time. %
We show the residual values for all posterior samples and for both frequency channels. %
In \cref{fig:selectedresiduals} we show residuals for three baselines on April 11th, similar to fig. 13 of \cite{ehtvi}.
Note that the apparent time evolution is largely due to the rotation of the Earth, and not due to intrinsic source variability.
Our inspection of the residuals validates that temporal changes in the data are captured by the reconstruction, as there is no systematic change of the residuals as time progresses for any of the four periods. %

By using only closure quantities, station-dependent calibration terms have been fully projected out for our reconstruction.
Since \cite{ehtiii} does not only perform partial calibration but also estimates the magnitude of the residual gains, performing self-calibration on our reconstruction provides an important consistency check.
Our reconstruction is not a single result but rather a collection of approximate posterior samples, so individual calibration solutions need to be computed for each of them.
Thereby, we obtain an uncertainty estimate on the gains, which we expect to be consistent with the pre-calibrated gains from the telescope.

\begin{figure*}
  \centering
  \includegraphics[width=\textwidth]{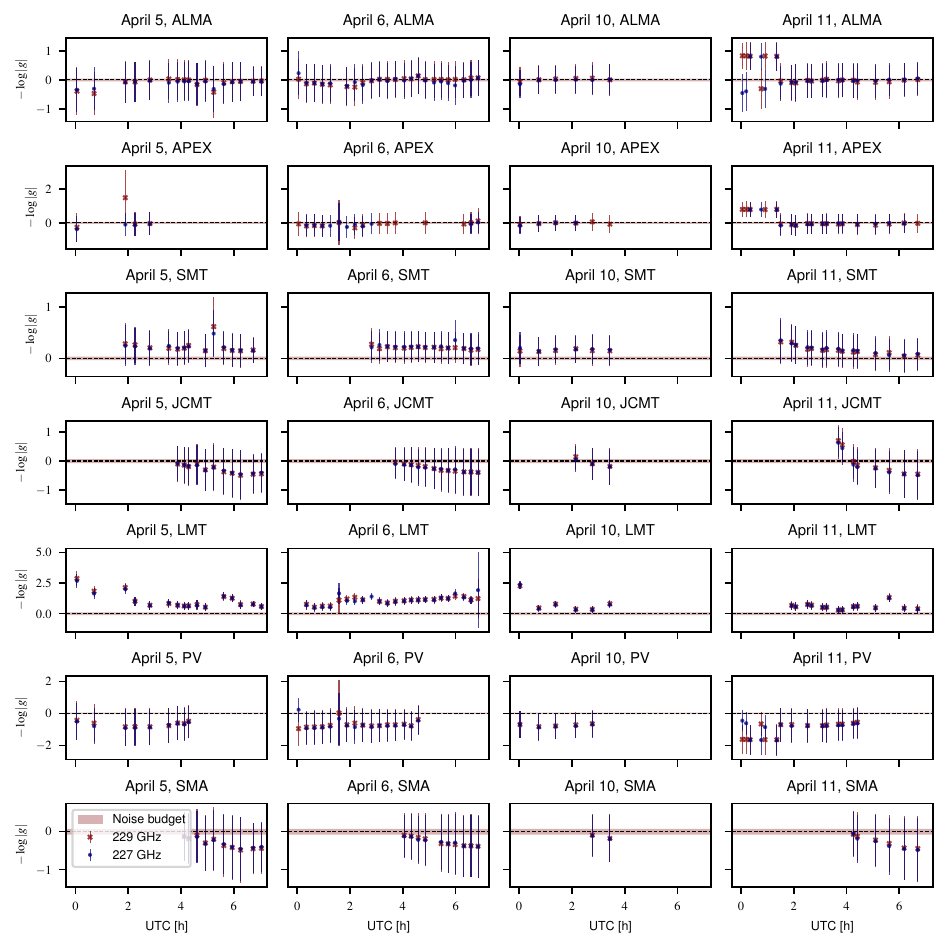}
  \caption[Normalized negative log-amplitude residual gains]{
    Normalized negative log-amplitude residual gains.  
    Mean and  1-$\sigma$ standard deviation of normalised negative log-amplitude residual gains $g$. The a priori noise budget is taken from \cite[tab.~14]{ehtiv}.}
  \label{fig:log_amplitudes}
\end{figure*}

\begin{figure*}
  \centering
  \includegraphics[width=\textwidth]{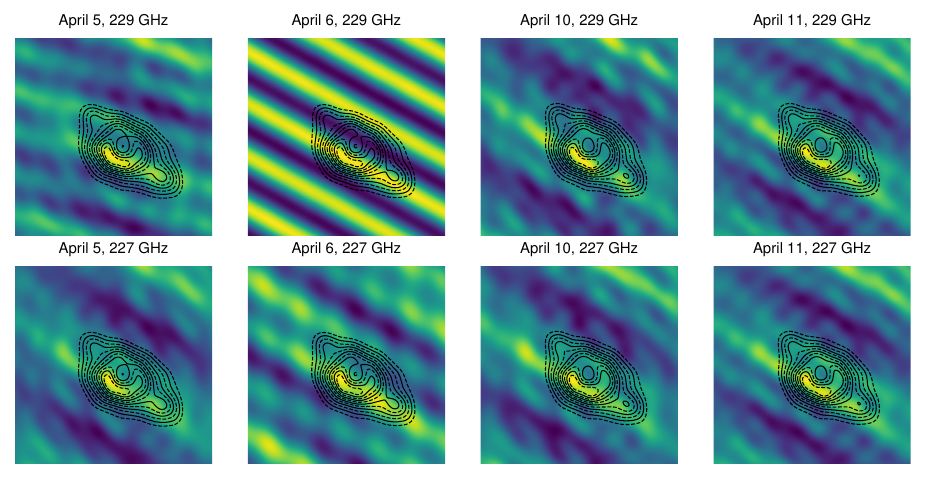}
  \caption[Sample-averaged dirty images of self-calibrated data]{Sample-averaged dirty images of self-calibrated data.
    The over-plotted contour lines show the brightness of the approximate posterior mean in multiplicative steps of $\nicefrac{1}{\sqrt{2}}$ and start at the maximum of the approximate posterior mean of our reconstruction.
    The solid lines correspond to factors of powers of two from the maximum.
  }
  \label{fig:selfcalibrated_dirty}
\end{figure*}

The negative log-amplitude-gains for all stations and days are shown in \cref{fig:log_amplitudes}, and, for reference, \cref{fig:selfcalibrated_dirty} depicts the sample-averaged dirty images of the calibrated data, overlaid with contours of the posterior mean image.
We can reproduce the issues with the calibration of the station LMT that have been reported by the EHT collaboration \cite{ehtiv}.
Apart from those, the pre-calibrated visibilities agree with our result within the uncertainty.

\subsection*{Reconstruction of the Correlation Structure}
\label{sec:powersp}
The recovered spatial correlation structures for the log-brightness, as well as the brightness itself are shown in \cref{fig:ps}. %
The relation between the power spectrum of the brightness $P_s$ and the log-brightness $|A|^2$ is given by:
\begin{equation}
	P_s \propto F e^{F^{-1} |A|^2} \ ,
\end{equation}
where $F$ denotes the Fourier transformation as defined in \cref{eq:skymodel}. %
On large scales, these agree with the ground truth to within the error bounds. %
Our examples do not have prominent small-scale features, so the ground truth power spectra drop off rapidly. %
We have only limited data on these scales due to the measurement setup, so the reconstruction is primarily informed by the prior distribution. %
As the prior favours power-law like behavior, the large scale information about the slope of the spectrum is extrapolated as a straight line towards small-scale modes. %
Therefore, deviations from a straight line cannot be captured in these regions and the variability of these deviations is limited by the prior variance. %

In addition, the posterior statistical properties of the power spectrum cannot be fully captured by the variational approximation of MGVI. \@
In particular, for small-scale features, the posterior uncertainty is asymmetric since deviations above and below the mean have an asymmetric effect on the observed data:
if the mean power of these scales is small compared to the power on large scales, further decreasing the power on these scales has almost no effect on the observed data whereas increasing the small-scale power has a significant impact. %
The forced symmetry of the posterior uncertainty can lead to an over-estimation of the small-scale power as the uncertainty towards less power is underestimated  (see \cref{fig:ps}).

On large image scales, where good constraints from the data are available, the correlation matches the ground truth exceptionally well, including characteristic features such as the disk diameter. %
The spectra of the two simulations based on the EHT imaging challenge, \texttt{challenge1} and \texttt{challenge2}, are an exception. %
We believe that the mismatch is explained by the diverse and pronounced structure of the simulations on all scales that cannot be resolved by the data. %

\section*{Data Availability}
The data this work is based on have been published by the Event Horizon Collaboration \cite{ehtiii, ehtiv} and are available at \cite{ehtdata}. %
We provide a set of 160~antithetic sample pairs of the sky brightness from the approximate posterior distribution, which can be used to propagate uncertainty to any derived quantity. %
The samples are available at \cite{samples}. %

\section*{Code availability}
The software sources used for producing the results of this publication are available at \cite{zenodo_software}.

\section*{Acknowledgements}
We thank Landman Bester and Iniyan Natarajan for discussions regarding VLBI imaging, %
the Schneefernerhaus for their hospitality,
and the five anonymous referees for numerous comments that significantly improved the manuscript, in particular for providing the prototype of \cref{fig:hierarchicalmodel}. %
P.A.\ acknowledges the financial support by the German Federal Ministry of Education and Research (BMBF) under grant 05A17PB1 (Verbundprojekt D-MeerKAT). %
J.K.\ acknowledges the financial support by the Excellence Cluster ORIGINS, which is funded by the Deutsche Forschungsgemeinschaft (DFG, German Research Foundation) under Germany's Excellence Strategy - EXC-2094-390783311.

\section*{Author Contributions}
All authors contributed text to this publication.
P.A., P.F., P.H., J.K and R.L.\  implemented and tested the instrument response, likelihood, and model.
J.K.\  developed the inference heuristic.
P.A.\ and J.K.\ performed the hyperparameter study.
P.F.\  and P.A.\  contributed the amplitude model which features outer products of power spectra. %
M.R.\  provided implementations and numerical optimisation for many of the employed algorithms. %
T.E.\  coordinated the team and contributed to discussions.

\section*{Competing interests}
The authors declare no competing interests.

\sloppy
\onecolumn{
\printbibliography
}
\fussy

\typeout{get arXiv to do 4 passes: Label(s) may have changed. Rerun}
\end{document}